\documentclass[twocolumns,10pt]{IEEEtran}
\usepackage[T1]{fontenc}%
\usepackage{caption,balance}
\usepackage{amssymb,amsmath,amsfonts,amsthm}
\usepackage{mathrsfs}
\usepackage{cite}
\usepackage{array}
\usepackage{bm}
\usepackage{tabularx}
\usepackage[table]{xcolor}
\usepackage{multicol, multirow}
\usepackage{color}
\usepackage{mathtools}
\usepackage{booktabs}
\usepackage{subcaption}
\usepackage{verbatim}
\usepackage{mathtools}
\usepackage{tikz,pgf}
\usetikzlibrary{
	calc,
	positioning,
	mindmap,
	trees,
	arrows.meta,
	fit,
	backgrounds,
	decorations.pathreplacing
}
\usepackage{graphicx}
\usepackage{bbm}
\usepackage[utf8]{inputenc}
\usepackage{pgfplots}
\usepackage{algorithm,algcompatible}

\newcommand{\blue}[1]{{\textcolor[rgb]{0,0,0}{#1}}}

\newtheorem{remark}{Remark}
\newtheorem{definition}{Definition}

\newtheorem{assumption}{Assumption}
\newtheorem{theorem}{Theorem}

\algblockdefx{FORALLP}{ENDFAP}[1]%
{\textbf{for all }#1 \textbf{do in parallel}}%
{\textbf{end for}}

\newenvironment{list4}{
	\begin{list}{$\bullet$}{%
			\setlength{\itemsep}{0.05cm}
			\setlength{\labelsep}{0.2cm}
			\setlength{\labelwidth}{0.3cm}
			\setlength{\parsep}{0in} 
			\setlength{\parskip}{0in}
			\setlength{\topsep}{0in} 
			\setlength{\partopsep}{0in}
			\setlength{\leftmargin}{0.16in}}}
	{\end{list}}

\usepackage[colorinlistoftodos,prependcaption,backgroundcolor=black!5!white,bordercolor=red]{todonotes}

\renewcommand{\L}{{\rm L}}
\newcommand{\A}{{\rm A}}
\newcommand{\sigcom}{\sigma_{\mathrm{com}}}
\newcommand{\sigsgd}{\sigma_{\mathrm{sg}}}
\newcommand{\D}{{\rm D}}
\newcommand{\W}{{\rm W}}
\newcommand{\J}{{\rm J}}
\newcommand{\I}{{\rm I}}
\newcommand{\N}{N}
\newcommand{\x}{x}
\newcommand{\y}{y}

\IEEEoverridecommandlockouts
\ifCLASSINFOpdf
\else
\fi
\usepackage[cmintegrals]{newtxmath}
\begin{document}
	\title{Distributed Consensus, Optimization, and Learning under Communication Constraints}
	
	\author{Zheng Chen, Nicolò Michelusi, Themistoklis Charalambous, and Kaibin Huang
		\thanks{Z. Chen is with the Dept. of Electrical Engineering, Linköping University, Linköping, Sweden (email: zheng.chen@liu.se). Her research is supported in part by ELLIIT, Swedish Research Council (VR), and Knut and Alice Wallenberg Foundation.} 
		\thanks{ N. Michelusi is with the School of Electrical, Computer and Energy Engineering, Arizona State University, AZ, USA (email: nicolo.michelusi@asu.edu). His research is funded by NSF CNS-2129015.}
		\thanks{ T. Charalambous is with the Dept. of Electrical and Computer Engineering, University of Cyprus, Cyprus (email: charalambous.themistoklis@ucy.ac.cy). His research is supported in part by the MINERVA project funded by the European Research Council (ERC) (Grant No. 101044629), and by the University of Cyprus with project acronym LOVELACE.}
		\thanks{K. Huang is with the Dept. of Electrical and Computer Engineering, The University of Hong Kong (HKU), Hong Kong (email: huangkb@hku.hk). His research is supported in part by the Research Grants Council of the Hong Kong Special Administrative Region, China under the NSFC/RGC CRS scheme $(\text{CRS}\_\text{HKU}702/24)$, the Areas of Excellence scheme grant $(\text{AoE}/\text{E-}601/22\text{-R})$ and the GRF Grants $17304925$ and $17212526$, and in part by the Croucher Senior Fellowship.}
	}
	
	\maketitle
	
	\begin{abstract}
		Many modern engineering applications require distributed problem-solving in which multiple nodes in a network cooperate to achieve a common optimization objective. Numerous distributed optimization and learning algorithms rely on consensus-based distributed coordination to ensure network-wide agreement via iterative local interactions. In these algorithms, the accuracy, frequency, and reliability of information exchange among network nodes directly influence system performance, including convergence behavior, training delay, and model accuracy. This tutorial article adopts a bottom-up approach to present the fundamentals of consensus averaging and distributed optimization methods, and highlights the impact of network connectivity, information flow, and communication uncertainty on network cooperation and optimization performance. Building on these theoretical insights, we discuss key design principles for communication-constrained distributed learning across diverse network topologies and review common approaches to communication-efficient learning, particularly in wireless networks.
		Despite substantial progress, further efforts are needed to bridge the gap between theory and practice and fully realize the potential of cooperative and networked intelligence.
	\end{abstract}
	
	
	\section{Introduction}
	Modern data-driven systems operate at increasingly large scales, with data generated by numerous edge devices distributed across wide geographical areas. Collecting all data at a central location for processing is often impractical due to communication constraints and privacy concerns. This motivates distributed algorithms that leverage local computational resources and information exchange to enable collaborative decision-making across the network \cite{10188495}. Distributed optimization provides a principled framework for collaboratively solving large-scale inference and decision-making problems using locally available information \cite{YANG2019278}. It has applications in a wide range of domains, including wireless sensor fusion, distributed estimation and inference \cite{6494683}, multi-agent coordination \cite{Nedic2018}, and collaborative machine learning (ML) \cite{verbraeken2020survey}.
	
	\begin{figure}[!t]
		\centering
		\includegraphics[width=.9\linewidth]{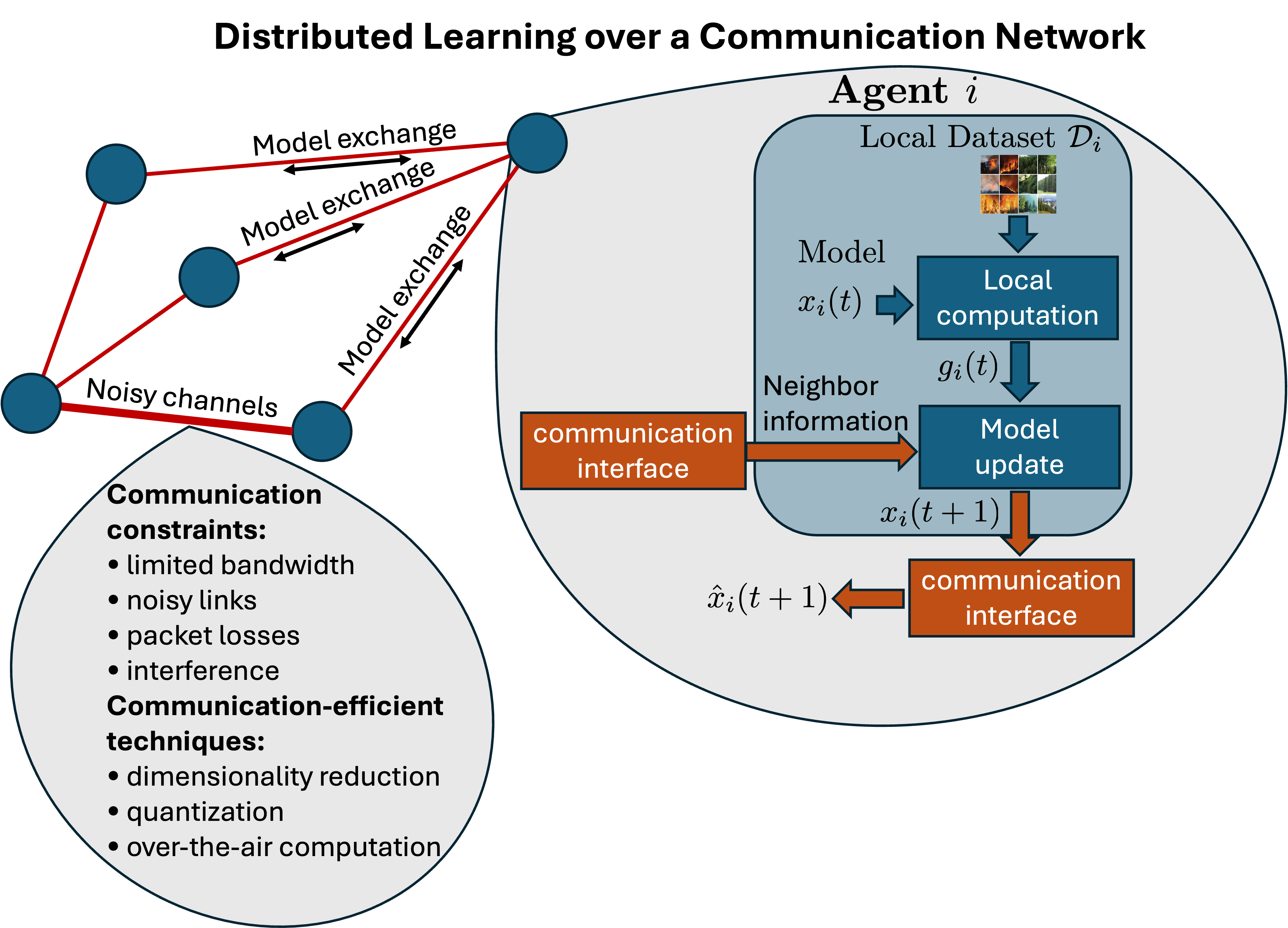}
		\caption{Illustration of a distributed learning network. Each agent performs local computation using its local dataset while exchanging model information with neighboring agents. The local model update combines locally computed information with models received from neighboring agents. }
		\label{fig:overview}
	\end{figure}
	
	\subsection{Historical Overview }
	A prominent class of distributed optimization and learning methods is based on consensus mechanisms. The study of distributed consensus has its roots in the classical problems of coordination and agreement in distributed systems and control engineering \cite{Tsitsiklis84,bertsekaspar}. The quantity of interest in consensus problems can be continuous- or discrete-valued. 
	For continuous-valued consensus, the seminal work of DeGroot \cite{degroot1974reaching} studies how a group reaches agreement on a common opinion (represented as a subjective probability distribution) through a linear update rule that iteratively computes weighted averages of the group members' opinions. By interpreting the weight matrix as the transition matrix of a Markov chain, the work establishes a fundamental connection between the convergence of the opinion dynamics and the properties of the underlying Markov chain. This line of research gained significant momentum with the rise of networked control, multi-agent systems, and sensor networks. A central concept in this area is average consensus, where a group of nodes/agents seeks to agree on the average of their initial states using only local information exchange.\footnote{Due to differences in terminology in control and ML communities, we use the terms node, agent, and client interchangeably throughout the article.} The seminal work \cite{xiao2004fast} formalizes average consensus as a linear iterative process and provides the necessary and sufficient conditions for convergence to the initial average. Several studies have revealed deep connections between consensus dynamics, spectral graph theory, and Markov chains \cite{4118472, olshevsky2009convergence}. Later on, average consensus has been extensively studied under more realistic conditions, including time-varying random graphs \cite{agreement-random,1431045,tahbaz2008necessary, 7194804}, directed links \cite{dominguez2011distributed,cai2012average,priolo2014distributed,hadjicostis2018distributed}, gossip communication \cite{kempe2003gossip, boyd2006randomized, 4497789, benezit2010weighted, dimakis2010gossip}, time delays \cite{1333204, bliman2008average, 6571230}, quantized updates \cite{carli2007average, 4542554, 9222206,11021366}, and link failures \cite{fagnani2009average, 5404434, 4663899, 10383677}. Beyond asymptotic convergence, a complementary line of work studies finite-time consensus that achieves exact agreement in a finite number of steps \cite{sundaram2007finite,  
		6760277, charalambous2015distributed, 8550443, 
		kibangou2012graph, 8619602}.
	
	The simplicity and theoretical explainability of average consensus made it a fundamental building block for distributed optimization and learning. Integrating first-order optimization methods using gradient or subgradient with consensus averaging has led to the emergence of decentralized gradient descent (DGD) \cite{2009:Nedic}. The basic principle is that each agent updates a local copy of the optimization variable by combining a gradient step on its local objective with a consensus step that incorporates information from neighbors. 
	A key insight was that consensus drives agreement among agents, while gradient steps drive optimality. Subsequent research extended DGD to more realistic settings, including stochastic gradients \cite{Nedic10}, time-varying topologies and directed links \cite{6930814}, asynchronous communication \cite{1104412, 9217472}, and quantized updates \cite{9782148, 9103200}. 
	
	Distributed optimization provides the algorithmic foundation for collaborative ML over decentralized data, where multiple agents jointly optimize a global objective expressed as the sum of empirical risk functions defined over their local datasets. 
	Depending on the network topology and system architecture, collaborative ML can be implemented in server-based, decentralized, or hierarchical (also called semi-decentralized) settings (see Fig.~\ref{fig:topology}). In a server-based architecture, such as Federated Learning (FL), a central parameter server (PS) aggregates local updates from distributed agents and updates the global model accordingly \cite{mcmahan2017communication}. In decentralized systems, agents communicate only with local neighbors and often employ consensus mechanisms to enforce network-wide agreement on model parameters \cite{lian2017can}. Hierarchical architectures \cite{9705093,semi-decen} combine intra-cluster aggregation among locally connected devices with inter-cluster aggregation assisted by higher-level servers/aggregators.
	Different network topologies and algorithmic designs introduce distinct communication flows within the network, through which information exchange contributes to the common goal of optimizing the global objective. 
	\vspace{-0.1cm}
	\subsection{Communication Constraints in Distributed Learning}
	The ability of distributed agents to exchange information, update their states, and coordinate their actions through local interactions or server-assisted synchronization is fundamental to achieving system-level cooperation. 
	The design of information mixing and the accuracy and frequency of communication directly influence the convergence, scalability, and robustness of distributed learning algorithms, particularly under data heterogeneity and resource constraints. As illustrated in Fig.~\ref{fig:overview}, the communication interface is an integral part of the information fusion process in distributed learning. In practical deployments, this interface operates over wireless networks subject to time-varying channel conditions, limited bandwidth and energy resources. 
	These communication constraints necessitate the joint design of communication protocols and resource allocation schemes to ensure efficient and reliable operation. Consequently, communication should be treated as an essential component of the algorithm design rather than a secondary implementation detail. 
	
	\vspace{-0.1cm}
	\subsection{Purpose and Organization}
	This tutorial provides a general introduction to the fundamentals of distributed consensus and optimization algorithms and underlines the connections between their convergence behavior and the design of information mixing mechanisms. The material complements existing tutorials on multi-agent consensus and cooperation \cite{8716798, 4118472}, first-order distributed optimization \cite{first-order}, and communication-efficient distributed learning \cite{9084352}.
	Our focus in this article is on characterizing the joint impact of network topology, information flow, and communication uncertainties (including client scheduling, link failures, and compression) on distributed learning performance in practical settings.
	
	The remainder of the paper is organized as follows. Sec.~\ref{sec:consensus} reviews the fundamentals of distributed linear consensus, including convergence conditions, weight design and optimization, the effects of imperfect communication, and ratio consensus over directed graphs. Sec.~\ref{sec:distributed_opt} presents classical distributed optimization algorithms, discusses steady-state errors and correction methods, and extends the discussion to collaborative ML with data sampling. Sec.~\ref{sec:comm_constraints} examines sources of communication uncertainty and their impact on distributed learning performance, while Sec.~\ref{numres} provides numerical illustrations of the theoretical results. Sec.~\ref{sec:comm-efficiency} reviews and categorizes communication-efficient methods. Finally, Sec.~\ref{sec:open} discusses open research questions and challenges, and Sec.~\ref{sec:conclusion} concludes the paper.
	
	\vspace{-0.1cm}
	\subsection{Notation}
	All vectors are column vectors. $\mathbf{1}$ is the vector of all ones (with dimension $N$).   $\J=\frac{1}{N}\mathbf{1}\mathbf{1}^{\top}$ is the orthogonal projection onto $\operatorname{span}\{\mathbf{1}\}$. $\I_d$ denotes the $d \times d$ identity matrix (or simply $\I$ when its dimension is clear from the context). $|\mathcal{S}|$ denotes the cardinality of a set $\mathcal{S}$. 
	For a vector $\mathbf{a}$, $\mathbf{a}^\top$ denotes its transpose 
	and
	$\|\mathbf{a}\|$ its Euclidean norm. 
	$\|A\|_2$ denotes the $2$-norm (or spectral norm) of the matrix $A$, and $\rho(A)$ its spectral radius, which are defined below.
	
	\begin{definition}
		Consider a matrix $A\in\mathbb{R}^{N\times N}$
		with eigenvalues $\lambda_1,\ldots, \lambda_N$. 
		The spectral radius of $A$ is defined as
		\begin{equation}
			\rho(A)=\max\limits_{i=1,\ldots,N}\{|\lambda_i|\}.
		\end{equation}
		The spectral norm of the matrix is \begin{equation}
			\|A\|_2=\max\limits_{\|x\|=1} \|Ax\|,
		\end{equation}
		which measures the maximum amplification of a vector by the matrix. 
		It is also equal to the largest singular value of $A$.
		These two quantities have the following relations:
		\begin{itemize}
			\item $\|A\|_2=\sqrt{\rho(A^\top A)}$;
			\item In general, $\rho(A)\leq \|A\|_2 $. When $A=A^\top$, $\rho(A)=\|A\|_2$. 
		\end{itemize}
	\end{definition}
	
	\section{Foundations of Distributed Consensus}
	\label{sec:consensus}
	In this section, we present: (1) fundamental concepts in graph theory and spectral graph analysis (Sec. \ref{basicgraph}); (2) key principles of distributed consensus, with a focus on average consensus implemented via linear iterative updates (Sec. \ref{basicconsensus} and \ref{sec:special-case}); (3) weight designs and convergence rate optimization (Sec. \ref{weight}); and (4) ratio consensus over directed graphs (Sec.~\ref{ratiocons}).
	Furthermore, we discuss how network connectivity and communication uncertainty influence the overall performance of consensus algorithms (Sec. \ref{consensuscomm}).
	
	\subsection{Basics on Graphs and Spectral Graph Theory}
	\label{basicgraph}
	A graph $\mathcal{G}=(\mathcal{V},\mathcal{E})$ consists of a set of nodes (or vertices) $\mathcal{V}=\{1,2,\ldots,N\}$ and a set of edges (or links) $\mathcal{E}\subseteq\mathcal{V}\times\mathcal{V}$. An edge $(i,j)\in\mathcal{E}$ represents a direct communication link from node $i$ to node $j$.
	For undirected graphs, edges have no direction, then $(i,j)\in \mathcal{E} $ implies $(j,i)\in \mathcal{E}, \forall i,j\in \mathcal{V}$, meaning that direct communication is possible in both directions.
	For directed graphs (or \textit{digraphs}), edges have directions, and $(i,j)\in\mathcal{E}$ indicates that node $i$ can directly transmit information to node $j$, but not necessarily vice versa. If $(i,j)\in\mathcal{E}$, then:
	\begin{itemize}
		\item node $j$ is an out-neighbor of node $i$;
		\item node $i$ is an in-neighbor of node $j$.
	\end{itemize}
	The in-neighbor and out-neighbor sets of each node $i\in\mathcal{V}$ are defined by $\mathcal{N}_i^{\rm{in}}=\{j\in \mathcal{V}| (j, i)\in \mathcal{E}\}$
	and $\mathcal{N}_i^{\rm{out}}=\{j\in \mathcal{V}| (i, j)\in \mathcal{E}\}$, respectively. The in-degree and out-degree of node $i$ are defined by $d_i^{\rm{in}}=|\mathcal{N}_{i}^{\rm{in}}|$ and $d_i^{\rm{out}}=|\mathcal{N}_{i}^{\rm{out}}|$, respectively. 
	For undirected graphs, the neighbor set is $\mathcal{N}_i=\mathcal{N}_i^{\rm{in}}=\mathcal{N}_i^{\rm{out}}$, and the \textit{degree} of node $i$ is $d_i=|\mathcal{N}_i|$. 
	
	\begin{definition}
		An undirected graph is \textit{connected} if there exists a path between any pair of nodes $i,j\in\mathcal{V}$.
		A directed graph is \textit{strongly connected} if there exists a path from every node $i\in\mathcal{V}$ to every other node $j\in\mathcal{V}$.
	\end{definition}
	
	Next, we introduce two important matrix representations of a graph: the adjacency matrix and the graph Laplacian.
	\begin{definition}
		The adjacency matrix of a graph $\mathcal{G}$ is the $N\times N$ square matrix $\A=[a_{ij}]$ given by
		\begin{equation}
			a_{ij}= 
			\left\lbrace 
			\begin{array}{ccc}
				1 & \text{if}~(j,i)\in \mathcal{E}, \\
				0	& \text{otherwise}.
			\end{array} \right.
		\end{equation} 
		For undirected graphs, $\A$ is symmetric, i.e., $\A^\top=\A$.
	\end{definition}
	In the remainder of this paper, we focus on \textit{undirected and connected} graphs unless otherwise specified.
	\begin{definition}
		The Laplacian matrix of an undirected graph is defined as $\L=\D-\A$,
		where $\D=\text{diag}(d_1,\ldots,d_N)$. Its elements are:
		\begin{equation}\nonumber
			\L_{i,j}= 
			\left\lbrace 
			\begin{array}{ccc}
				d_i &  &\text{if}~i=j, \\
				-1	&  &\text{if}~(j,i)\in \mathcal{E}, \\
				0 & &\text{otherwise}.
			\end{array} \right.
		\end{equation}
	\end{definition}
	It is easy to see that $\L\mathbf{1}=\mathbf{0}$, i.e., $0$ is an eigenvalue  of $\L$ with corresponding eigenvector $\mathbf{1}$. Let $\Delta=\max\limits_{i=1,\ldots, N}\{d_i\}$ be the maximum node degree. According to the Gershgorin circle theorem, the eigenvalues of $\L$ are located within $[0,2\Delta]$. Furthermore, if the graph is \textit{connected}, 
	then
	$0$ is a simple eigenvalue of $\L$, and all other eigenvalues are strictly positive. 
	
	\subsection{ Average Consensus over Time-Invariant Graphs}
	\label{basicconsensus}
	Reaching a consensus refers to the process of reaching an agreement on a certain quantity of interest among a group of nodes. Let $x_i(0)$ represent the initial state value of node $i$ and let $\bm x(0)=[x_1(0),x_2(0),\ldots, x_N(0)]^{\top}$
	be the initial state value vector. Consensus is reached when the state vector converges to a point in the consensus subspace, i.e., the one-dimensional subspace $\mathcal{C} = \operatorname{span}\{\mathbf{1}\}
	= \{\, c\,\mathbf{1} : c \in \mathbb{R} \,\}$. In particular, the case where all state values converge to the initial network average 
	$\bar{x} =\frac{1}{N}\sum_{i=1}^{N}x_i(0)$ is referred to as the \textit{average consensus}.
	
	The most fundamental algorithm for achieving average consensus is the following iterative update rule
	\begin{equation}
		x_i(t+1)=w_{ii} x_i(t)+\sum_{j\in\mathcal{N}_i}w_{ij}x_j(t),
		\label{eq:scalar_form}
	\end{equation}  
	where $w_{ij}$ is the weight that node $i$ assigns to the information (state value) received from node $j$ and $w_{ii}$ is the self weight.
	This means that a node 
	updates its state for the next iteration by averaging its own state value and the neighbors' state values in the current iteration. Note that $w_{ij}=0$ if $(j,i)\notin\mathcal{E}$ and $i\neq j$.
	Putting all node states in a vector, this linear iteration can be written in matrix form as
	\begin{equation}
		\bm x(t+1)=\W\bm x(t),
		\label{eq:linear_consensus_iteration}
	\end{equation}
	where $\W=[w_{ij}]_{i,j\in\mathcal{V}}$.
	This matrix is often referred to as the \textit{mixing matrix}, or \textit{weight matrix}, or \textit{combination matrix}.
	To reach average consensus asymptotically, all nodes' state values need to converge to the average of initial values $\bar{x}=\frac{1}{N}\mathbf{1}^{\top}\bm x(0)$.
	This means
	\begin{equation}
		\lim\limits_{t\rightarrow\infty}\W^{t}=\frac{1}{N}\mathbf{1}\mathbf{1}^{\top} \triangleq \J. 
		\label{eq:convergence}
	\end{equation}
	According to the analysis in \cite{xiao2004fast}, \eqref{eq:convergence} is achieved if the following spectral conditions on $\W$ are satisfied.
	\begin{assumption}[Spectral Conditions on Mixing Matrix]
		\label{assumptions:W}
		The mixing matrix $W$ is assumed to satisfy the following conditions:
		\begin{equation}
			\W\mathbf{1}=\mathbf{1},\quad \mathbf{1}^{\top}\W=\mathbf{1}^{\top}, \quad \rho\left(\W-\J\right)<1,
		\end{equation}
		where $\rho(\cdot)$ is the spectral radius.
	\end{assumption}
	The first condition implies that every vector in the consensus subspace is a fixed point of the mapping described in \eqref{eq:linear_consensus_iteration}.
	The second condition ensures that every consensus step preserves the average value (or the total mass) across the network, i.e.,  $\mathbf{1}^\top \bm x(t+1) =  \mathbf{1}^\top \bm x(t)$. Let $\lambda_i(\W), i=1\ldots N$, denote the eigenvalues of $\W$. Obviously, we have $\lambda_1(W)=1$. The last condition imposes the requirement that all other eigenvalues lie strictly within the unit circle. This implicitly requires that the graph is connected. Under Assumption~\ref{assumptions:W}, we have $\rho\left(\W-\J\right)=\max\limits_{i\geq 2}\{|\lambda_i(\W)|\}$, which is the second largest eigenvalue modulus of $\W$, sometimes denoted as $\lambda_2^{+}$.
	
	\begin{remark}
		An undirected graph (with a symmetric adjacency matrix) does not necessarily imply a symmetric mixing matrix. The spectral conditions in Assumption \ref{assumptions:W} do not require $\W$ to be non-negative or symmetric to ensure convergence to average consensus. However, assuming non-negativity and symmetry simplifies the convergence analysis of distributed consensus and optimization algorithms, which will be discussed later.
	\end{remark}
	
	The ideal consensus state is given by
	$\bar{\bm x}=\bar{x}\mathbf{1}=\J\bm x(0)$,
	which is the orthogonal projection of the initial state
	onto the consensus subspace
	$\mathcal{C}=\operatorname{span}\{\mathbf{1}\}$.
	To characterize convergence, we define the disagreement
	vector as
	$\boldsymbol{\delta}(t)=(\I-\J)\bm x(t)$,
	which satisfies
	$\mathbf{1}^\top\boldsymbol{\delta}(t)=0$.
	Since the network average is preserved,
	$\boldsymbol{\delta}(t)=\bm x(t)-\bar{\bm x}$.
	For a fixed mixing matrix $\W$ satisfying Assumption~\ref{assumptions:W},
	the disagreement evolves according to
	\begin{equation}
		\boldsymbol{\delta}(t+1)
		=(\W-\J)\boldsymbol{\delta}(t).
		\label{eq:disagreement}
	\end{equation}
	Consequently, asymptotic average consensus is achieved
	for any initial state if and only if
	$\rho(\W-\J)<1$.
	
	The convergence speed of the distributed linear iteration in \eqref{eq:linear_consensus_iteration} can be characterized by the following two metrics \cite{xiao2004fast}:
	\begin{itemize}
		\item asymptotic convergence factor
		\begin{equation}
			\label{asymconvfac}R_{\textnormal{asym}}(\W)=\sup\limits_{\bm x(0)\neq \bar{\bm x}}\lim\limits_{t\rightarrow \infty}\left(\frac{\|\bm x(t)-\bar{\bm x}\|}{\|\bm x(0)-\bar{\bm x}\|}\right)^{1/t}=\rho\left(\W-\J\right);
		\end{equation}
		\item per-step convergence factor
		\begin{equation}
			\label{perstepconv}R_{\textnormal{step}}(\W)=\sup\limits_{\bm x(t)\neq \bar{\bm x}}\frac{\|\bm x(t+1)-\bar{\bm x}\|}{\|\bm x(t)-\bar{\bm x}\|}=\big\|\W-\J\big\|_2.
		\end{equation}
	\end{itemize}
	
	\subsection{Doubly Stochastic Matrix and Perron-Frobenius Theorem}
	\label{sec:special-case}
	
	In a special case where $\W$ is primitive (hence, non-negative, see Definition \ref{nonnegmatrix} and discussions thereafter), 
	the dominant eigenvalue is real, positive, and simple. Then 
	the limit of $\W^t$ can be analyzed using the Perron-Frobenius theorem.
	
	\begin{definition}
		\label{nonnegmatrix}
		A square matrix $A\in\mathbb{R}^{N\times N}$ is non-negative if all entries $A_{i,j}\geq 0, \forall i, j$. 
	\end{definition}
	An important property of non-negative matrices is that the spectral radius $\rho(A)$ is an eigenvalue. A non-negative matrix is \textit{irreducible} if its associated directed graph is strongly connected. Then, by Perron–Frobenius theorem \cite[Th. 8.4.4]{matrixanalysis}, $\rho(A)$ is a simple eigenvalue (with algebraic multiplicity $1$). An irreducible non-negative matrix is \textit{primitive} if it has only one eigenvalue with maximum modulus (i.e., $\lambda_1=\rho(A)$), and all other eigenvalues satisfy $|\lambda_i|<\rho(A), i=2,\ldots,N$. 
	\begin{theorem}\cite[Th. 8.5.1]{matrixanalysis}
		\label{theorem:perron}
		If a non-negative matrix $A\in\mathbb{R}^{N\times N}$ is primitive, with left and right eigenvectors $\mathbf{u}$ and $\mathbf{v}$ associated to the Perron eigenvalue $\lambda_1=\rho(A)$, respectively, satisfying \[
		A\mathbf{v}=\lambda_1\mathbf{v}, \qquad \mathbf{u}^\top A=\lambda_1\mathbf{u}^\top,\] normalized such that $\mathbf{v}^\top \mathbf{u}=1$, then $\lim\limits_{t\rightarrow\infty }A^t/\lambda_1^t=\mathbf{v}\mathbf{u}^\top$.
	\end{theorem}
	
	\begin{definition}
		\label{def:A}
		A non-negative matrix $A\in \mathbb{R}^{N\times N}$ is called 
		\begin{itemize}
			\item row-stochastic if  $A\mathbf{1}=\mathbf{1}$;
			\item column-stochastic if  $\mathbf{1}^\top A=\mathbf{1}^\top$;
			\item doubly stochastic if  $A\mathbf{1}=\mathbf{1}$ and  $\mathbf{1}^\top A=\mathbf{1}^\top$.
		\end{itemize}
	\end{definition}
	Furthermore, if the mixing matrix $\W$ is doubly stochastic and primitive, then $\rho(\W)=1$ is its Perron eigenvalue, and $\mathbf{u}=\mathbf{v}=\frac{1}{\sqrt{N}}\mathbf{1}$ are the corresponding (normalized) left and right Perron eigenvectors. It  follows directly that $\lim\limits_{t\rightarrow\infty }\W^t=\frac{1}{N}\mathbf{1}\mathbf{1}^\top$. Then, average consensus is achieved asymptotically.
	
	\begin{remark}
		Not all doubly stochastic matrices with a strongly connected communication graph are primitive. An example is $\W=[0~1; 1~0]$, which swaps the state values of two connected nodes. $\W$ has two eigenvalues ($1$ and $-1$) with maximum modulus. Obviously, $\lim\limits_{t\rightarrow\infty }\W^t$ does not exist.
	\end{remark}
	
	\begin{remark}
		A doubly stochastic mixing matrix satisfying the conditions in Assumption \ref{assumptions:W} can still be asymmetric even when the underlying communication graph is undirected. 
		However, numerous studies still rely on the symmetric mixing assumption, in which case all eigenvalues are real-valued, and the matrix is orthogonally diagonalizable.
		For this reason, in Sec.~\ref{sec:distributed_opt}, we primarily focus on doubly stochastic and symmetric mixing matrices.
	\end{remark}
	
	\subsection{Mixing Matrix Design}
	\label{weight}
	
	In this subsection, we review some commonly adopted mixing matrix designs and show how the mixing weights can be optimized to achieve faster convergence.
	
	Two commonly considered mixing matrix designs are:
	\begin{enumerate}
		\item Constant edge weights: $$w_{ij}=\beta, \forall(j,i)\in\mathcal{E},$$ where $\beta$ is chosen such that $0<\beta<1/\Delta$, with $\Delta=\max\limits_{i=1,\ldots, N}\{d_i\}$. 
		\item Local-degree weights (Metropolis): $$w_{ij}=\frac{1}{1+\max\{d_i,d_j\}}, (j,i)\in\mathcal{E}.$$
	\end{enumerate}
	In both cases, the self-weights are chosen as $w_{ii}=1-\sum_{j\in\mathcal{N}_i}w_{ij}$, thus the mixing matrix $\W$ is doubly stochastic.
	In particular, for the case with constant edge weights, we have 
	\begin{equation}
		\W=\I-\beta \L.
	\end{equation}
	As a result, the $i$-th eigenvalue (ranked in decreasing order) of $\W$ is $\lambda_i(\W)=1-\beta \lambda_i(\L)$, where $\lambda_i(\L)$ is the $i$-th eigenvalue of graph Laplacian $\L$, ranked in increasing order. It is easy to see that this design of $\W$ satisfies all the conditions in Assumption~\ref{assumptions:W} when the graph is connected.
	
	In the general case, the mixing matrix can be optimized by solving the following optimization problem, aiming to minimize the asymptotic convergence factor \eqref{asymconvfac}:
	\begin{subequations}
		\begin{eqnarray}
			&\textnormal{minimize}  &\rho\left(\W - \J\right)\\
			&\textnormal{subject~to}& \W\mathbf{1}=\mathbf{1}, \mathbf{1}^\top \W=\mathbf{1}^\top,\label{12b} \\
			&&\W \in \mathcal{S}.\label{12c}
		\end{eqnarray}
	\end{subequations}
	$\mathcal{S}$ imposes a sparsity pattern that is consistent with the underlying communication graph, i.e., $\mathcal{S}=\{\W\in\mathbb{R}^{N\times N}\big|[\W]_{i,j}=0 \textnormal{~if~} (j,i)\notin \mathcal{E}, \textnormal{~and~} i\neq j \}$.
	However, the spectral radius is a non-convex objective in general.  Since $\rho(\A)\leq \|\A\|_2$, an alternative consideration is to minimize the spectral norm (equivalent to minimizing the per-step convergence factor \eqref{perstepconv}):
	\[
	\textnormal{minimize}\ \|\W - \J\|_2,\ 
	\textnormal{subject~to}\ \eqref{12b}-\eqref{12c}.
	\]
	This can be transformed into a semi-definite programming (SDP) problem \cite{xiao2004fast}, solvable using standard convex optimization tools:
	\begin{subequations}
		\begin{eqnarray}
			&\textnormal{minimize}  &s\\
			&\textnormal{subject~to}& \begin{bmatrix}
				s\I & \W - \mathbf{1}  \mathbf{1}^{\top} / N \\
				\W^\top - \mathbf{1}  \mathbf{1}^{\top} / N & s\I
			\end{bmatrix}\succeq 0 \label{eq:constraint}\\
			&& \eqref{12b}-\eqref{12c}.
		\end{eqnarray}
	\end{subequations}
	When $\W$ is symmetric, \eqref{eq:constraint} becomes $-s \I\preceq \W - \mathbf{1}  \mathbf{1}^{\top} / N \preceq s \I$.
	
	\subsection{Consensus with Communication Uncertainty}
	\label{consensuscomm}
	So far, we have focused on deterministic settings, where every node communicates with its neighbors at every iteration and all transmitted messages are received without error. We now briefly discuss extensions to time-varying mixing matrices and noisy communications.
	\subsubsection{Time-Varying Mixing}
	The mixing matrix $\W(t)$ may vary over time for several reasons:
	\begin{itemize}
		\item network topology changes over time (due to, for instance, node mobility and time-varying channel conditions);
		\item communication decisions change over time, as in asynchronous communication frameworks, such as randomized gossip or other forms of communication scheduling;
		\item transmission outcomes change over time, due to packet losses or link failures. 
	\end{itemize}
	In the first two cases, the stochasticity conditions on the mixing matrix can be enforced, and the convergence behavior can be analyzed by studying the infinite product of random stochastic matrices and its connections to the ergodicity of non-homogeneous Markov chains \cite{product-stochastic-matrices, tahbaz2008necessary, agreement-random}. However, in the last case, 
	additional mechanisms need to be applied to compensate for the information loss and ensure convergence \cite{fagnani2009average,7219386,10383677}. 
	
	\subsubsection{Consensus with Noisy Communication}
	\label{sec:consensus-noisy}
	When nodes exchange their state values, the communication can be noisy, for instance, due to either additive noise in the channel or quantization noise in the compression process. 
	In the presence of additive communication noise, the consensus iteration can be expressed as
	\begin{equation} x_i(t+1)=w_{ii}x_i(t)+\sum\nolimits_{j\in\mathcal{N}_i}w_{ij}x_j(t)+e_i(t),
	\end{equation}
	where $e_i(t)$ denotes the effective noise at node $i$, which may represent either node-specific receiver noise or the aggregate contribution of noise from its incoming communication links.
	Under the assumption that the additive noise is independent and identically distributed (i.i.d.) across nodes and iterations, with zero mean and finite variance, it was shown in \cite{xiao2007distributed} that the mean-square deviation from the instantaneous network average converges to a finite steady-state value, even though the network average itself evolves as an unbounded random walk.
	For link-level additive noise, subsequent studies have established that stochastic approximation algorithms with diminishing stepsizes can achieve almost sure consensus, with all node states converging to a common random limit \cite{4663899, huang2009coordination}. 
	This limiting value is unbiased only in expectation, and it generally differs from the exact average in any single realization due to the accumulated noise over time.
	Furthermore, a leakage-based algorithm was introduced in \cite{leakage} to mitigate this random bias by incorporating the initial value into the consensus updating rule, and a similar mechanism was adopted in \cite{noise-resilient} for directed graphs. Recently, \cite{CDC-yuhang} reformulated average consensus as a distributed optimization problem and derived a noise-resilient linear recursion from a noisy DGD perspective, providing a unified interpretation of several existing approaches.
	
	Another source of communication error is quantization. Let $\mathcal{Q}$ denote a quantization operator, with the quantization error at node $i$ defined as $e_i(t)=\mathcal{Q}(x_i(t))-x_i(t)$. For deterministic quantizers, the error generally depends on the local state and may introduce systematic bias. In contrast, randomized or dithered quantizers can be designed to produce conditionally zero-mean errors. Stochastic approximation algorithms with diminishing stepsizes can mitigate the effects of unbiased quantization errors, whereas persistent quantization bias may require additional compensation mechanisms.
	
	\subsection{Ratio Consensus over Directed Graphs}
	\label{ratiocons}
	Common consensus algorithms developed for undirected graphs rely on the symmetry of information flows between nodes. If an undirected graph is connected, we can construct a doubly stochastic and primitive mixing matrix that is compatible with the graph topology. This ensures that the total mass (or average) of the node states is preserved and that all nodes asymptotically converge to the exact average of their initial values.
	However, in directed graphs, communication is generally asymmetric, and computing a doubly stochastic mixing matrix using only local information is challenging and may require additional weight-balancing mechanisms; see, for example,~\cite{gharesifard2012distributed,rikos2014distributed,charalambous2014minimum,9222206} and references therein.
	As a result, standard average consensus algorithms no longer preserve the total mass, and the algorithm may converge to a biased value or fail to converge. One approach to overcome this limitation is \textit{ratio consensus} (or \textit{push sum})  by augmenting the state at each node with an additional variable and applying the same column-stochastic mixing weights to both states, so that the ratio of the two variables converges to the desired average \cite{kempe2003gossip, dominguez2010coordination, 6571230, rezaeinia2019push}. A more thorough discussion of distributed averaging (including ratio consensus) and balancing in directed graphs can be found in~\cite{hadjicostis2018distributed}.
	
	\section{Foundations of Distributed Optimization}
	\label{sec:distributed_opt}
	
	Distributed optimization aims to cooperatively minimize a global objective expressed as an aggregate of local functions across distributed nodes/agents. It is a key enabler of modern large-scale ML, particularly in federated and decentralized settings where data are distributed across heterogeneous agents.
	Consider a network of $\N$ agents connected through a communication graph $\mathcal{G}=(\mathcal{V},\mathcal{E})$, where each agent $i\in\mathcal{V}$ holds a private local objective function $f_i:\mathbb{R}^d\rightarrow\mathbb{R}$. The problem of interest is
	\begin{equation}
		\x^\star \in \arg\min_{\x \in \mathbb{R}^d} F(\x), \quad F(\x)\triangleq\frac{1}{\N}\sum\nolimits_{i=1}^\N f_i(\x).
		\label{eq:global_objective}
	\end{equation}
	Since the local objective functions are not shared, each agent performs local computations based on its own objective and exchanges local estimates of the optimization variables with its neighbors (Fig.~\ref{fig:overview}). Consensus mechanisms coordinate these local estimates, enabling the agents to iteratively approach the global optimizer in a fully distributed manner.
	
	In this section, we introduce the algorithmic foundations of distributed optimization, starting with DGD and its stochastic counterpart (DSGD). We then discuss steady-state bias and representative techniques for achieving exact convergence, followed by several extensions beyond the classical DGD/DSGD framework. Finally, we connect distributed optimization to distributed learning using a unified cooperative SGD framework for different communication architectures.
	
	First, we impose a set of standard assumptions.
	\begin{assumption}[Network Connectivity]
		\label{assump:connectivity}
		The communication graph $\mathcal{G}=(\mathcal{V},\mathcal{E})$ is
		connected.
	\end{assumption}
	
	\begin{assumption}[Mixing Matrix]
		\label{assump:mixing}
		The mixing matrix $\W$ is symmetric ($\W^\top=\W$), doubly stochastic, with $\rho\left(\W-\J\right)<1$.
	\end{assumption}
	A symmetric matrix has real-valued eigenvalues. Under this assumption, the eigenvalues of $\W$ can be sorted as $1=\lambda_1>\lambda_2\geq\dots \geq \lambda_N>-1$.
	
	\begin{assumption}[Strong Convexity]
		\label{assump:strong_convexity}
		Each local objective function $f_i:\mathbb{R}^d \to \mathbb{R}$ is
		$\mu$-strongly convex, i.e.,
		\vspace{-0.15cm}
		\begin{equation}
			f_i(\y)\geq f_i(\x) +
			\nabla f_i(\x)^\top (\y-\x) + \frac{\mu}{2}\|\x-\y\|^2, \forall \x,\y \in \mathbb{R}^d,
		\end{equation}
		for some constant $\mu>0$.
	\end{assumption}
	Strong convexity ensures that the global objective
	$F(\x)$ admits a \emph{unique} minimizer
	$\x^\star$. This property is crucial for establishing linear
	convergence rates for distributed algorithms with constant stepsizes.
	In the absence of strong convexity, convergence guarantees typically weaken to sublinear rates or stationarity conditions.
	\begin{assumption}[Smoothness]
		\label{assump:smoothness}
		Each local objective function $f_i$ is $L$-smooth, i.e., its gradient is
		Lipschitz continuous with parameter $L$:
		\vspace{-0.15cm}
		\begin{equation}
			\|\nabla f_i(\x)-\nabla f_i(\y)\|
			\;\le\;
			L\|\x-\y\|,
			\qquad \forall \x,\y \in \mathbb{R}^d.
		\end{equation}
	\end{assumption}
	Smoothness limits how rapidly the gradient can change with respect to the optimization variables, making it a key assumption in the convergence analysis of distributed first-order methods.
	
	These assumptions represent the classical setting of distributed optimization over undirected graphs with simple consensus-based algorithms such as DGD and DSGD. Extensions to directed communication graphs require modified consensus mechanisms, as discussed in Sec.~\ref{beyondDGD}.
	
	\subsection{Distributed Gradient Descent (DGD)}
	\label{sec:DGD}
	The classical DGD algorithm generalizes gradient descent to multi-agent networks via a consensus–optimization iteration. Each agent $i$ maintains a local variable $\x_i(t)$ updated by
	\begin{align}
		\x_i(t+1)=\sum\nolimits_{j=1}^N w_{ij} \x_j(t)-\alpha_t \nabla f_i\left(\x_i(t)\right),
		\label{eq:DGD}
	\end{align}
	where $\W=[w_{ij}]_{i,j\in[N]}$ is the mixing matrix, $\nabla f_i(x_i(t))$ is the gradient of the local objective function evaluated at the current iterate, and $\alpha_t$ is the stepsize (or learning rate). 
	Since each agent maintains its local copy $x_i(t)$ of the decision variable and computes gradients using its own local objective, the local parameters generally differ across the network. 
	By stacking the local parameters over the network, we can rewrite it as 
	\begin{equation}
		\label{noiselessDGD}
		\bm x(t+1) = (\W \otimes \I_d) \bm x(t) - \alpha_t \nabla \mathbf{f}(\bm x(t)),
	\end{equation}
	where $\bm x(t) = [x_1(t)^\top, \ldots, x_\N(t)^\top]^\top \in \mathbb{R}^{\N d}$ collects all local parameters in one vector, and $\nabla \mathbf{f}(\bm x(t))=[\nabla f_1\left(\x_1(t)\right)^\top, \ldots, \nabla f_N\left(\x_N(t)\right)^\top]^\top$ collects all local gradient vectors in the same order.
	Under Assumptions~\ref{assump:connectivity},~\ref{assump:mixing},~\ref{assump:strong_convexity}, and~\ref{assump:smoothness},
	with constant stepsize $\alpha_t=\alpha \leq (1+\lambda_N)/(\mu+L)$, 
	the squared error (SE) of DGD is bounded as \cite{10947567}
	\begin{equation}
		\frac{1}{\N}\sum_{i=1}^\N \|x_i(t){-}x^\star\|^2
		{\leq} (1{-}\alpha\mu)^{2t}\frac{2}{\N}\sum_{i=1}^\N \|x_i(0) {-}x^\star\|^2
		{+}
		8\frac{\alpha^2 \kappa^2\Delta b}{(1{-}\lambda_2)^2},
	\end{equation}
	where $\kappa = L/\mu$ is the condition number
	and $\Delta b \triangleq \frac{1}{\N}\sum_{i=1}^\N \|\nabla f_i(\x^\star)\|^2$ measures gradient heterogeneity at the global optimum, i.e., local and global optima do not coincide. Therefore, DGD converges linearly to a neighborhood of the global optimum $\x^\star$, with contraction factor $(1-\alpha\mu)^2$. This limiting neighborhood is the \emph{steady-state bias} caused by persistent disagreement across the network, and its size depends on the stepsize $\alpha$, the spectral properties of $\W$ (second largest eigenvalue $\lambda_2$), and gradient heterogeneity $\Delta b$.\footnote{
		Here, the factor
		$1/(1-\lambda_2)=\|(\I-\W)^\dagger\|_2$
		characterizes the network dependence of the DGD
		steady-state error bound. One should not confuse $\lambda_2$
		with the consensus contraction factor
		$\rho(\W-\J)=\max\{|\lambda_2|,|\lambda_N|\}$.}
	
	On the other hand, with a properly chosen diminishing stepsize sequence $\alpha_t=\mathcal O(1/t)$,
	the iterates converge to the exact optimizer, with the SE satisfying
	\begin{equation}
		\label{DGDconv}
		\frac{1}{\N}\sum\nolimits_{i=1}^\N \|x_i(t) - x^\star\|^2
		= \mathcal{O}(1/t^2).   
	\end{equation}
	
	Note that there are variants of decentralized gradient-based methods other than \eqref{noiselessDGD}, such as
	\begin{enumerate}
		\item Adapt-then-Combine (ATC) diffusion:\\
		$\bm x(t+1) = (\W \otimes \I_d)\big[\bm x(t) - \alpha_t\, \nabla \mathbf{f}(\bm x(t))\big]$;
		\item Combine-then-Adapt (CTA) diffusion:\\
		$\bm x(t+\tfrac12) = (\W \otimes \I_d) \bm x(t)$,\\     
		$\bm x(t+1) = \bm x(t+\tfrac12) - \alpha_t\, \nabla \mathbf{f}(\bm x(t+\tfrac12))$.
	\end{enumerate}
	The main differences lie in whether the consensus operation is applied before or after the gradient step, and at which model the gradient is evaluated. As a result, these methods require different stability and convergence conditions. Nevertheless, all of them suffer
	from a steady-state bias under constant stepsizes, caused by gradient dissimilarity.
	
	\subsection{Distributed Stochastic Gradient Descent (DSGD)}
	\label{sec:DSGD}
	
	In DSGD, at each iteration $t$, agent $i$ does not have access to the exact gradient $\nabla f_i(x_i(t))$ at its current local iterate $x_i(t)$, but instead observes a noisy first-order oracle
	\[
	g_i(t) = \nabla f_i(x_i(t)) + \zeta_i(t),
	\]
	where $\zeta_i(t)$ is a random perturbation process capturing uncertainty in the gradient information. 
	Such uncertainty arises from, for example, mini-batch sampling or streaming data in online learning settings (see more detailed discussions in Sec. \ref{opt2learn}). 
	The DSGD update rule is then
	\begin{equation}
		\x_i(t+1)=\sum_{j=1}^\N w_{ij} \x_j(t)-\alpha_t g_i(t),
		\label{eq:DSGD}
	\end{equation}
	which replaces the exact gradient in
	\eqref{eq:DGD} with its stochastic approximation.
	In compact vector form,
	\begin{equation}
		\bm x(t+1)=(\W \otimes \I_d)\bm x(t)-\alpha_t \mathbf{g}(t),
		\label{dsgd}
	\end{equation}
	where $\mathbf{g}(t)$ denotes the stacked local gradient estimators.
	In addition to the steady-state bias inherent to DGD, stochastic gradient noise introduces a second source of asymptotic error. Hence, rather than converging deterministically to the limiting neighborhood, the iterates exhibit persistent random fluctuations around it. As a result, the steady-state mean-square error (MSE) increases, yielding a larger asymptotic error than in the deterministic setting (see \eqref{eq:DSGD_ss}).
	
	To study the convergence of DSGD,
	we introduce some additional assumptions.
	\begin{assumption}
		\label{assump:sfo}
		The stochastic gradient $g_i(t)$ is a conditionally unbiased estimator of the exact gradient $\nabla f_i(x_i(t))$, with finite variance $\sigsgd^2$. That is,
		\begin{subequations}
			\begin{align}
				&\mathbb{E}\left[g_i(t)\mid \bm x(t)\right]=\nabla f_i(x_i(t)),\\
				&\mathbb{E}\left[\|g_i(t)-\nabla f_i(x_i(t))\|^2 \mid \bm x(t)\right]\le\sigsgd^2.
				\label{ineq:grad_noise}
			\end{align}   
		\end{subequations}
	\end{assumption}
	
	Under Assumptions~\ref{assump:connectivity}-\ref{assump:sfo},
	and a sufficiently small constant stepsize $\alpha_t = \alpha$,
	DSGD converges linearly to a neighborhood of the global minimizer $\x^\star$, with steady-state error bounded as \cite[Sec. V.A]{10947567}:
	\begin{equation}
		\limsup_{t \to \infty} \mathbb{E}\left[\frac{1}{\N}\sum_{i=1}^\N \|x_i(t) - x^\star\|^2\right] \leq 2\frac{\alpha\sigsgd^2}{\mu}+2 \frac{\alpha^2 \kappa^2 \Delta b}{(1-\lambda_2)^2}.
		\label{eq:DSGD_ss}
	\end{equation}
	The first term in \eqref{eq:DSGD_ss} reflects the impact of stochastic gradient noise and matches the steady-state behavior of centralized SGD, while the second term is a heterogeneity-induced bias, observed also in noiseless DGD. 
	This asymptotic error floor can be mitigated by using diminishing stepsizes, $\alpha_t =\mathcal O(1/t)$. In this case,
	DSGD achieves MSE
	convergence to $x^\star$ at a sublinear rate of $\mathcal{O}(1/t)$, requiring $\mathcal{O}(1/\epsilon)$ iterations to reach $\epsilon$-accuracy \cite{9609587,8619228}. 
	This rate is slower than DGD with exact gradient computation, due to the impact of gradient noise.
	
	These observations highlight several fundamental challenges of DGD/DSGD:
	\begin{list4}
		\item \textit{Objective Heterogeneity:}
		When the local objective functions differ, gradient heterogeneity may introduce a persistent optimization bias, resulting in a nonzero error floor under a constant stepsize and preventing exact convergence to the global optimum.
		
		\item \textit{Gradient Noise:}
		Stochastic gradients introduce persistent random fluctuations around the limiting neighborhood, increasing the steady-state error (under a constant stepsize), or slowing convergence ($\mathcal{O}(1/t)$ vs $\mathcal{O}(1/t^2)$ with diminishing stepsizes).
		
		\item \textit{Network Dependence:}
		The asymptotic error bound scales as $(1-\lambda_2)^{-2}$, indicating greater sensitivity to gradient heterogeneity in poorly connected networks (e.g., sparse or high-diameter graphs). 
		
	\end{list4}
	
	\subsection{Gradient Tracking and Variance Reduction}
	To overcome the limitations of DGD/DSGD, modern distributed learning methods augment the basic framework with two key techniques, each targeting the two main sources of steady-state error of DSGD: gradient heterogeneity ($\Delta b$) and stochastic gradient variance ($\sigsgd^2$).
	
	\subsubsection{Gradient Tracking (GT)}
	Each agent $i$ holds a local estimate $y_i(t)$ of the global average gradient
	$
	\frac{1}{\N} \sum_{j=1}^\N \nabla f_j( x_j(t))$,
	evaluated
	at the agents’ current local iterates \cite{2016:NEXT,2017:NedicOlshevsky,2019:Usman,2021:PuNedic}.
	Using this estimate, agent $i$ updates its local model according to
	\begin{equation}
		x_i(t+1) = \sum_{j=1}^\N w_{ij} x_j(t) - \alpha_t y_i(t),
	\end{equation}
	where the term $-\alpha_t y_i(t)$ plays the role of a gradient descent step with respect to the global objective function.
	
	After the model update is performed, each agent updates its gradient-tracking variable as
	\begin{equation}
		y_i(t+1) = \sum_{j=1}^\N w_{ij} y_j(t)
		+ \nabla f_i\bigl(x_i(t+1)\bigr)
		- \nabla f_i\bigl(x_i(t)\bigr).
	\end{equation}
	This recursion combines consensus on local gradient estimates with a correction term that compensates for the variation in local gradients caused by the model updates.\footnote{For simplicity, we focus on the case where local gradient computations are noiseless, although the framework can be extended to noisy gradients in a manner similar to DSGD.} Initializing $y_i(0)=\nabla f_i\bigl(x_i(0)$, the recursion preserves $\frac{1}{N}\sum_{i=1}^{N}y_i(t)=\frac{1}{N}\sum_{i=1}^{N}\nabla f_i\bigl(x_i(t)\bigr)$. 
	
	With GT, the local gradient $\nabla f_i(x_i(t))$ in DGD is replaced by a tracked global gradient estimate $y_i(t)$. By employing dynamic average consensus to track time-varying local gradients, GT estimates the global gradient at each agent without centralized coordination. This decouples consensus from optimization and eliminates the bias term $\Delta b$, enabling exact convergence to $x^\star$ under noiseless updates despite gradient heterogeneity. However, this improvement requires additional communication overhead, as agents must exchange both local model variables and global gradient estimates.
	
	\subsubsection{Variance Reduction (VR)}
	Persistent gradient noise of SGD can be reduced by constructing unbiased estimators whose variance diminishes as the iterates approach the global optimum. To conceptually illustrate the idea, we focus on SAGA~\cite{defazio2014saga} in a server-based architecture, with stochasticity arising from random client/agent selection while assuming exact local gradient computations. More specifically,
	the PS maintains a table of past local gradients, with one entry per agent, which is updated incrementally over time.
	At iteration $t$, a single agent index $i_t \in \{1,\dots,\N\}$ is selected uniformly at random, and the SAGA gradient estimator is formed as
	\begin{equation}
		\textstyle{\tilde g(t) = \nabla f_{i_t}(x(t))-\nabla f_{i_t}(\tilde x_{i_t})
			+\frac{1}{\N}\sum_{j=1}^{\N} \nabla f_j(\tilde x_j),}
	\end{equation}
	where $\nabla f_j(\tilde x_j)$ denotes the gradient stored in memory for agent $j$, evaluated at its most recent reference point $\tilde x_j$. After the gradient estimator is formed, the memory table is updated by refreshing only the entry corresponding to the selected agent $i_t$. Specifically, the reference point and stored gradient are updated as $\tilde x_{i_t} \gets x(t)$, and
	$\nabla f_{i_t}(\tilde x_{i_t}) \gets \nabla f_{i_t}(x(t))$, respectively, while all other memory entries remain unchanged.  
	It is easy to see that $\tilde g(t)$ is an unbiased estimate of the global gradient, i.e.,
	taking the expectation with respect to the client selection, $\mathbb E\left[\tilde g(t)|x(t),\{\tilde{x}_j\}_{j=1}^{\N}\right]
	=\nabla F(x(t))$. Moreover, as $x(t),\tilde x_j\to x^\star,\forall j$, the variance of $\tilde g(t)$ vanishes, and hence  $\tilde g(t)\to\nabla F(x^\star)= 0$.
	
	Another related method is Stochastic Variance Reduced Gradient (SVRG)~\cite{NIPS2013_ac1dd209}, which uses a periodically updated snapshot point and its full gradient as a control variate. Unlike SAGA, SVRG avoids storing one gradient per agent but requires periodic full-gradient aggregation. Extensions of both methods can also incorporate stochastic or mini-batch local gradients, thereby effectively reducing variance due to both client and data sampling.
	
	\begin{remark}
		For both SVRG and SAGA, the variance of the gradient estimator vanishes as the algorithm converges to the optimum. Consequently, for smooth and strongly convex objectives, these variance reduction methods achieve linear convergence with constant stepsizes, comparable to deterministic gradient descent, while maintaining a low per-iteration computational cost through stochastic gradient evaluations.
	\end{remark}
	
	\subsubsection{Variance Reduction for Graph-Structured Networks}
	Variance reduction methods do not directly extend to graph-structured networks. 
	For example, with SVRG,
	exact computation of the global gradient at the snapshot point would require running a consensus procedure to convergence, which incurs significant communication overhead and undermines the
	communication efficiency of SVRG.
	This limitation has motivated the development of decentralized variance-reduction methods that replace exact global gradient computation with gradient tracking (or dynamic consensus mechanisms), enabling variance reduction while preserving localized communication and fully distributed operation.
	Combining gradient tracking with variance reduction yields algorithms such as 
	GT-SAGA and GT-SVRG \cite{9226112},
	which operate on undirected and weight-balanced directed
	graphs.
	Under the assumption of a primitive and doubly stochastic mixing matrix $\W$ (not necessarily symmetric), they
	achieve:
	\begin{list4}
		\item \textit{Exact Linear Convergence:} For strongly-convex and smooth objectives, these methods converge linearly to the exact optimum $x^\star$ with \emph{constant} stepsizes.
		\item \textit{Sample Complexity:} The total number of local gradient computations per agent to achieve a target level $\epsilon$ of MSE accuracy  is
		$\mathcal{O}(\max\{M,\kappa^2(1-
		\|\W-\J\|_2)^{-2}\} \log(1/\epsilon))$, where $M$ is the number of data samples per agent \cite[Theorem 1]{9226112}.
	\end{list4}
	
	\begin{table*}[ht]
		\centering
		\caption{Representative distributed first-order optimization algorithms and their convergence properties, under the assumptions of strongly convex and smooth local objectives.
		}
		\label{tab:distributed_algorithms}
		\renewcommand{\arraystretch}{1.15}
		\begin{tabular}{lcccccc}
			\toprule
			\textbf{Algorithm}
			& \textbf{Mixing}
			& \textbf{Noise Source}
			& \textbf{Stepsize}
			& \textbf{SE/MSE Conv. Rate}
			& \textbf{Optimality}
			& \textbf{Reference}\!\! \\
			\midrule
			
			DGD
			& Doubly stoch.
			& ---
			& Constant
			& Linear ($\mathcal{O}(c^t)$)
			& Neighborhood
			& \cite{10947567} \\
			
			DGD
			& Doubly stoch.
			& ---
			& Diminishing
			& $\mathcal{O}(1/t^2)$
			& Exact
			& \cite{10947567} \\
			
			DSGD
			& Doubly stoch.
			& Gradient
			& Diminishing
			& $\mathcal{O}(1/t)$
			& Exact
			& \cite{9609587} \\
			
			DSGD
			& Doubly stoch.
			& \!\!\!\!\!\!Random links/gradient\!\!\!\!\!
			& Diminishing
			& $\mathcal{O}(1/t)$
			& Exact
			& \cite{8619228} \\
			
			DSGD
			& Doubly stoch.
			& \!\!\!\!\!Comm. noise/gradient\!\!\!\!\!
			& Diminishing
			& $\mathcal{O}(1/\sqrt{t})$
			& Exact
			& \cite{10680589} \\
			
			DIGing (DGD-GT)\!\!\!\!
			& Doubly stoch.
			& \!\!\!\!\!Time-varying network\!\!\!\!\!
			& Constant
			& Linear 
			& Exact
			& \cite{2017:NedicOlshevsky} \\
			
			GT-SVRG/SAGA\!\!\!\!
			& Doubly stoch.
			& Gradient
			& Constant
			& Linear 
			& Exact 
			& \cite{9226112} \\
			
			EXTRA
			& Doubly stoch.
			& ---
			& Constant
			& Linear 
			& Exact
			& \cite{shi2015extra} \\
			
			Push-DIGing
			& Column stoch.
			& \!\!\!\!\!Time-varying network\!\!\!\!\!
			& Constant
			& Linear 
			& Exact
			& \cite{2017:NedicOlshevsky} \\
			
			AB/Push-Pull
			& Row-/Column-stoch.
			& ---
			& Constant
			& Linear 
			& Exact
			& \cite{pushpull2021} \\
			
			\bottomrule
		\end{tabular}
	\end{table*}
	
	\subsection{Beyond the Classic DGD/DSGD Framework}
	\label{beyondDGD}
	Practical distributed learning departs from the classical DGD/DSGD framework through several important relaxations, including time-varying topologies, directed information flows, and asynchronous operation. In what follows, we briefly review representative extensions along these directions.
	\begin{list4}
		\item \textbf{Time-varying topologies:} 
		In practice, the communication graph often evolves over time due to agent mobility, intermittent participation, link failures, or dynamic scheduling policies. Although the resulting communication patterns vary across iterations, convergence can still be established under suitable connectivity assumptions~\cite{6930814,koloskova2020unified}.
		
		\item \textbf{Directed communication links:} Communication networks may exhibit asymmetric information flows, due to heterogeneous transmit powers or unidirectional communication links. In such settings, the mixing matrix may no longer be symmetric or doubly stochastic.
		Directed-network algorithms rely on additional mechanisms, such as push-sum normalization (see Sec.~\ref{ratiocons})~\cite{benezit2010weighted}, weight-balancing techniques~\cite{9222206}, or separate row- and column-stochastic mixing matrices.
		Under suitable connectivity assumptions, these methods still achieve exact convergence~\cite{AB2018,pushpull2021}.
		
		\item 
		\textbf{Asynchronous operation:} Heterogeneous computation capabilities and communication delays often prevent synchronous operation. Asynchronous algorithms allow agents to compute and update independently without waiting for all others to complete each iteration. However, stale information and delayed updates complicate the convergence analysis and may slow convergence. Representative approaches include~\cite{recht2011hogwild,cederberg2025an}.
		
	\end{list4}
	
	Table~\ref{tab:distributed_algorithms} summarizes classical DGD/DSGD algorithms and the representative extensions, highlighting their mixing strategies, noise models, stepsize design, convergence rates, and optimality guarantees.
	
	\subsection{From Optimization to Learning}
	\label{opt2learn}
	The most prominent application of distributed optimization is collaborative training of ML models over decentralized data. In such settings, the global objective is often expressed as the empirical risk evaluated over the ensemble of all local datasets. Consider a supervised ML model $p_{\x}:\mathcal{F}\rightarrow\mathcal{Y}$ parameterized by a vector $\x\in\mathbb R^d$ that maps an input feature vector $\phi\in \mathcal{F}$ to an output label $y\in\mathcal{Y}$. Given a dataset of $M_{\mathrm{tot}}$ datapoints (input-output pairs) $\mathcal{D}=\{(\phi_m,y_m)\}_{m=1,\ldots,M_{\mathrm{tot}}}$, and the loss function $\ell:\mathcal{Y}\times \mathcal{Y}\rightarrow \mathbb{R}$, the empirical risk over the dataset $\mathcal D$ is defined as
	\begin{equation}
		\mathcal{R}(\x; \mathcal{D})=\frac{1}{M_{\mathrm{tot}}}\sum\nolimits_{m=1}^{M_{\mathrm{tot}}}\ell(y_m,p_{\x}(\phi_m)).
		\label{globER}
	\end{equation}
	The goal of model training is to find the parameter vector $x^\star$ that minimizes the empirical risk:
	\begin{align}
		x^\star=\arg\min_{x\in\mathbb R^d}
		\ \mathcal{R}(\x; \mathcal{D}).
		\label{ERmin}
	\end{align}
	In a distributed setting with $\N$ agents, each agent holds a local dataset $\mathcal{D}_i=\{(\phi^i_m,y^i_m)\}_{m=1,\ldots,M_i}$ with $M_i$ data samples.
	Thus, the global dataset in the entire network is
	$\mathcal{D}=\cup_{i=1}^\N
	\mathcal{D}_i$.
	The local objective can be written as $f_i(\x)=\mathcal{R}(\x; \mathcal{D}_i)$, and the global objective takes the form
	\begin{equation}
		F(\x)=\sum\nolimits_{i=1}^{N}\frac{M_i}{M_{\mathrm{tot}}}f_i(\x),
	\end{equation}
	where $M_{\mathrm{tot}}=\sum_{i=1}^{N}M_i$ is the total number of training samples in the system. Note that $F(\x)$ is equivalent to
	the empirical risk $\mathcal{R}(\x; \mathcal{D})$ over the global dataset, defined in \eqref{globER}.
	
	A key distinction with respect to the formulation in \eqref{eq:global_objective} is that the contribution of agent $i$ is weighted proportionally to its local dataset size, $M_i/M_{\mathrm{tot}}$, rather than uniformly by $1/\N$.
	This mismatch can be readily addressed by rescaling the local objective functions $f_i(\cdot)$ by a factor of $\N M_i / M_{\mathrm{tot}}$.
	For simplicity of exposition, we henceforth assume that all agents hold datasets of equal size, i.e., $M_i = M$ for all $i$.
	The empirical risk minimization problem in \eqref{ERmin} can then be addressed using the algorithmic frameworks presented in Sec. \ref{sec:distributed_opt}, across
	all distributed learning architectures depicted in Fig. \ref{fig:topology}.
	
	In large-scale ML with a massive amount of training data ($N\cdot M\gg 1$), mini-batch sampling is often applied to reduce the cost of gradient computation.
	At each iteration, each
	agent computes its local gradient using only a single datapoint or
	a small mini-batch, rather than the full local dataset.
	This data sampling process introduces noise in the gradient computation. Specifically, in the local training step, each agent $i$ obtains a stochastic gradient under the current local model $x_i$ as
	\begin{equation}
		g_i=\nabla \mathcal{R}(\x_i; \xi_i),
	\end{equation}
	where $\xi_i\subseteq \mathcal{D}_i$ is the randomly sampled mini-batch with $B_i$ samples. 
	When these $B_i$ samples are selected uniformly at random from the local dataset $\mathcal D_i$ (without replacement), it is straightforward to see that
	$$
	\mathbb E[g_i|\x_i]=\nabla f_i(x_i),
	$$
	i.e., the stochastic gradient $g_i$ is an unbiased estimate of the local gradient. Furthermore, its variance
	$\mathrm{var}(g_i|\x_i)\triangleq
	\mathbb E[\Vert g_i-\nabla f_i(x_i)\Vert^2|\x_i]$
	can be expressed as
	\cite[Prop. 1]{wu2020noisy}
	$$
	\mathrm{var}(g_i|\x_i)
	=
	\frac{M_i-B_i}{B_i(M_i-1)}
	\mathrm{var}_{(\phi,y)\sim\mathcal U(\mathcal D_i)}(\nabla\ell(y,p_{\x_i}(\phi))),
	$$
	where 
	$\nabla\ell(y,p_{\x_i}(\phi))$ is the gradient of the loss function (with respect to the local parameter vector $x_i$) and
	$\mathrm{var}_{(\phi,y)\sim\mathcal U(\mathcal D_i)}(\nabla\ell(y,p_{\x_i}(\phi)))$ is its variance, with respect to a datapoint $(\phi,y)$ selected uniformly at random from the local dataset $\mathcal D_i$.
	When the loss function $\ell(\cdot)$ is $L$-smooth with respect to the parameter vector $\x$, we can further bound \cite[Appendix A]{10680589}
	\begin{equation}
		\mathrm{var}(g_i|\x_i)
		\leq
		\frac{M_i-B_i}{B_i(M_i-1)}
		\Big[
		\sigma_{\mathrm{sg}\star}
		+ L\Vert \x_i-\x^\star\Vert\Big]^2,
		\label{eq:variance-bound}
	\end{equation}
	where $\sigma_{\mathrm{sg}\star}^2\triangleq
	\max_{i\in\mathcal{V}}\{\mathrm{var}_{(\phi,y)\sim\mathcal U(\mathcal D_i)}(\nabla\ell(y,p_{\x^\star}(\phi)))\}$
	denotes the maximum variance of the sample-wise gradients evaluated at the optimal solution $\x^\star$ in \eqref{ERmin}, across all agents.
	
	Notably, in the regime of small mini-batch sizes ($B_i \ll M_i$), the variance of the mini-batch gradient estimator is inversely proportional to the mini-batch size as $\mathcal{O}(1/B_i)$. 
	Unlike the uniform bounded-variance assumption in Assumption \ref{assump:sfo}, the upper bound in \eqref{eq:variance-bound} depends on the distance to the global optimum and may grow as the iterates move away from it.
	This additional error term may lead to instability if not properly controlled, but can be mitigated by choosing a sufficiently small stepsize \cite{10947567}.
	
	\subsection{Unified Framework on Cooperative SGD}
	Collaborative ML algorithms under different network structures (e.g., fully decentralized, semi-decentralized, and server-based) can be unified within a common theoretical framework. Fundamentally, they are all special cases of distributed optimization where the mixing matrices can be either fixed or time-varying to capture the effects of communication dynamics and model aggregation/mixing.
	
	\subsubsection{Distributed Synchronous SGD}This setting consists of a central PS and multiple clients that synchronize their models through server-assisted aggregation. In each communication round, all clients participate by performing one local SGD step on their local data before the server aggregates the updated models. The mathematical representation of this algorithm is equivalent to ATC diffusion with  $\W=\frac{1}{\N}\mathbf{1}\mathbf{1}^\top$. 
	
	\subsubsection{Federated Averaging (FedAvg)}The main distinctive features of FedAvg compared with distributed synchronous SGD are \emph{partial client participation} and \emph{multiple local SGD updates} between communication rounds. Suppose that one communication round takes place every $\tau$ local SGD steps and that, in each communication round, $K$ out of $N$ clients are selected randomly. Then the local models from selected clients are averaged at the server, and the resulting new model is broadcast to all clients. An equivalent matrix representation is obtained by defining the time-varying mixing matrix
	\[
	\W(t)=
	\begin{cases}
		\dfrac{1}{K}\mathbf{1}\mathbf{s}(t)^\top, & t \equiv 0 \pmod{\tau},\\[1mm]
		\I, & t \not\equiv 0 \pmod{\tau},
	\end{cases}
	\]
	where $\mathbf{s}(t)\in\{0,1\}^N$ denotes the client-selection vector. Using this mixing matrix representation, it is readily verified that $\W(t)\mathbf{1}=\mathbf{1}$, whereas $\mathbf{1}^{\top}\W(t)\neq\mathbf{1}^{\top}$ in general. Thus, although consensus can be achieved, the consensus value is generally biased toward clients with higher participation probabilities \cite{cho2022towards}. If all clients have the same participation probability, then $\mathbf{1}^{\top}\mathbb{E}[\W(t)]=\mathbf{1}^{\top}$, and convergence in expectation to the exact minimizer $x^\star$ can still be achieved under suitable assumptions.
	\subsubsection{Semi-Decentralized Learning}
	Suppose that inter-cluster aggregation occurs every $\tau$ iterations. The mixing matrix alternates between a sparse matrix $\W^{(1)}$, representing intra-cluster communication with peer-to-peer information exchange, and an aggregation matrix $\W^{(2)}$, representing periodic inter-cluster synchronization coordinated through the server.
	
	\section{Communication Constraints in Distributed Optimization and Learning}
	\label{sec:comm_constraints}
	
	In distributed learning systems, each node typically computes local model updates based on its own local data, but convergence to the global optimum requires consistent communication and coordination with other nodes in an iterative process. 
	In this section, we will review the following aspects: (1) the sources of communication uncertainty in distributed learning over modern wireless networks (Sec.~\ref{sec:comm-uncertainty}); (2) information flows and communication costs under different system architectures (Sec.~\ref{sec:comm-architecture}); (3) latency and reliability trade-offs in DSGD, and the impact of communication noise on convergence (Sec.~\ref{sec:dsgd-noisy}).
	
	\subsection{Communication Uncertainty in Distributed Learning}
	\label{sec:comm-uncertainty}
	\paragraph{Link Reliability}
	Modern data networks are designed using a layered architecture, in which each layer performs specific functions that enable reliable communication between a source and a destination. 
	Packet transmission from one point to another may fail due to random channel variations, contention for the shared communication medium, and hardware impairments.
	When information transmission occurs over wireless channels, the probability of packet failure (or outage) depends on many factors, such as the transmit power, noise and interference levels, and time-varying channel conditions. 
	Although higher-layer protocols implement mechanisms such as retransmission and error control to improve reliability, these mechanisms cannot fully eliminate physical-layer randomness under stringent latency, bandwidth, or energy constraints. 
	Consequently, distributed learning and inference systems operate under communication uncertainty: communication links act as stochastic operators whose random realizations directly affect the information exchanged among agents and, consequently, the collective system performance.
	
	\paragraph{Data Compression}
	Local model updates, such as gradients or model parameters, are inherently continuous-valued and must be represented using a finite number of bits before transmission over digital communication links. Since communication resources are limited, the available communication budget often does not permit transmitting high-precision updates. For example, in an Orthogonal Frequency Division Multiplexing (OFDM) system, each user is allocated only a limited number of resource blocks. Consequently, model updates are typically compressed prior to transmission.
	The maximum number of bits that can be reliably conveyed over a communication link depends on both the allocated communication resources and the channel conditions, which together determine the achievable compression ratio. The resulting compression process (including quantization and source coding) introduces distortion and information loss as structured noise in the received updates, affecting both consensus and optimization dynamics.
	
	\paragraph{Intermittent Node Participation}
	
	In addition to link-level impairments, distributed learning systems often face uncertainty arising from intermittent node participation: nodes may temporarily disconnect from the network due to client scheduling, energy constraints, mobility, or a lack of incentives to participate continuously \cite{wang2022unified}. As a result, the set of active participating nodes may vary over time, leading to time-varying network topologies and information flows. This intermittent participation introduces additional randomness into the learning process, affecting consensus formation, information mixing, and convergence behavior. 
	
	Overall, communication uncertainty arising from link failures, quantization, compression, and intermittent node participation
	fundamentally shapes the performance of distributed learning algorithms, by slowing down information mixing, biasing model updates, and introducing steady-state errors. 
	This motivates the development of distributed optimization and learning frameworks that are both communication-aware and resilient to highly dynamic and uncertain environments.
	
	\subsection{Communication Architectures}
	\label{sec:comm-architecture}
	
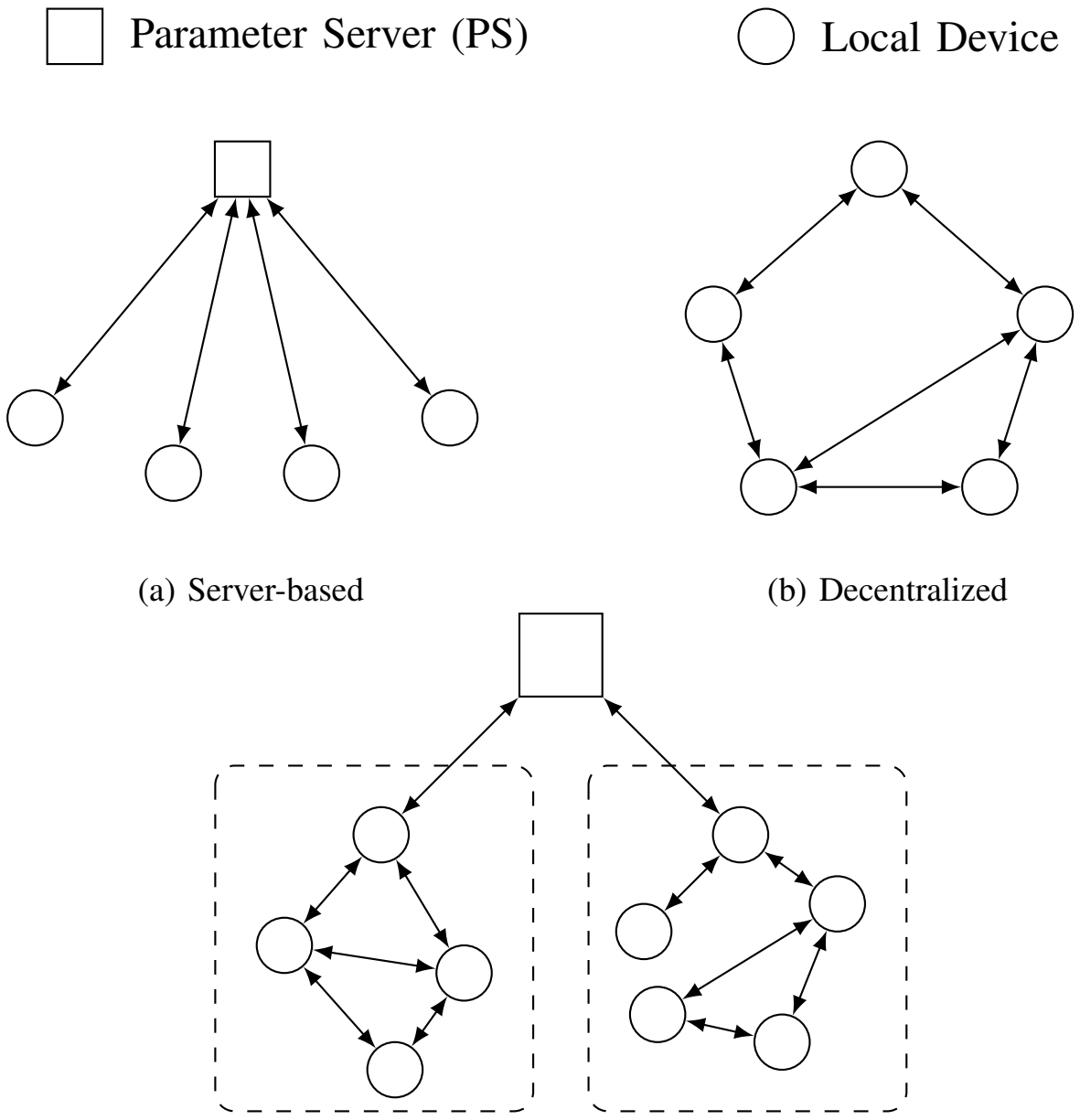
\begin{figure}[t]
		\centering
		\begin{tikzpicture}[node/.style={draw, circle, minimum size=0.4cm}, 
			server/.style={draw, rectangle, minimum size=0.4cm}]
			\node[server] (ps_legend) at (-1.5,2) {};
			\node[right=2pt of ps_legend] {\small Parameter Server (PS)};
			
			\node[node] (dev_legend) at (3.5,2) {};
			\node[right=2pt of dev_legend] {\small Local Device};
		\end{tikzpicture}
		\vspace{0.5cm}
		
		\begin{subfigure}[t]{0.48\linewidth}
			\centering
			\begin{tikzpicture}[
				server/.style={draw, rectangle, minimum size=0.4cm},
				worker/.style={draw, circle, minimum size=0.4cm},
				>=latex
				]
				\node[server] (ps) at (0,1.8) {};
				\node[worker] (w1) at (-1.5,-0.5+0.5) {};
				\node[worker] (w2) at (-0.5,-0.9+0.5) {};
				\node[worker] (w3) at (0.5,-0.9+0.5) {};
				\node[worker] (w4) at (1.5,-0.5+0.5) {};
				\foreach \w in {w1,w2,w3,w4}{
					\draw[<->] (ps) -- (\w);
				}
				\vphantom{\node at (0,-2+1) {};}
			\end{tikzpicture}
			\vspace{-0.2cm}
			\caption{Server-based}
		\end{subfigure}
		\hfill
		\begin{subfigure}[t]{0.48\linewidth}
			\centering
			\begin{tikzpicture}[
				node/.style={draw, circle, minimum size=0.4cm},
				>=latex
				]
				\node[node] (n1) at (0,1.8) {};
				\node[node] (n2) at (1.2,0.5+0.25) {};
				\node[node] (n3) at (0.8,-1+0.5) {};
				\node[node] (n4) at (-0.8,-1+0.5) {};
				\node[node] (n5) at (-1.2,0.5+0.25) {};
				\draw[<->] (n1)--(n2);
				\draw[<->] (n2)--(n3);
				\draw[<->] (n3)--(n4);
				\draw[<->] (n4)--(n5);
				\draw[<->] (n5)--(n1);
				\draw[<->] (n2)--(n4);
				\vphantom{\node at (0,-2+1) {};}
			\end{tikzpicture}
			\vspace{-0.2cm}
			\caption{Decentralized}
		\end{subfigure}
		\begin{subfigure}[t]{0.6\linewidth}
			\centering
			\begin{tikzpicture}[
				server/.style={draw, rectangle, minimum size=0.6cm},
				node/.style={draw, circle, minimum size=0.4cm},
				>=latex
				]
				\node[server] (ps) at (0,1.8) {};
				
				\node[node] (c1a) at (-1.3,0.5) {};
				\node[node] (c1b) at (-2,-0.3) {};
				\node[node] (c1c) at (-0.7,-0.5) {};
				\node[node] (c1d) at (-1.2,-1.2) {};
				\draw[<->] (c1a)--(c1b);
				\draw[<->] (c1a)--(c1c);
				\draw[<->] (c1b)--(c1c);
				\draw[<->] (c1b)--(c1d);
				\draw[<->] (c1c)--(c1d);
				\draw[<->] (ps)--(c1a);
				
				\node[node] (c2a) at (1.3,0.5) {};
				\node[node] (c2b) at (2,0) {};
				\node[node] (c2c) at (1.6,-1) {};
				\node[node] (c2d) at (0.7,-0.8) {};
				\node[node] (c2e) at (0.6,-0.2) {};
				\draw[<->] (c2a)--(c2b);
				\draw[<->] (c2a)--(c2e);
				\draw[<->] (c2b)--(c2c);
				\draw[<->] (c2b)--(c2d);
				\draw[<->] (c2d)--(c2c);
				\draw[<->] (ps)--(c2a);
				
				\draw[dashed, rounded corners] (-2.5,1) rectangle (-0.2,-1.5);
				\draw[dashed, rounded corners] (0.2,1) rectangle (2.5,-1.5);
				
				\vphantom{\node at (0,-2) {};}
			\end{tikzpicture}
			\vspace{-0.2cm}
			\caption{Semi-decentralized}
		\end{subfigure}
		
		\caption{Illustration of common distributed learning architectures.}
		\label{fig:topology}
	\end{figure}

	Depending on how nodes cooperate and on the underlying system architecture, various challenges arise in optimizing resource allocation under communication constraints. Here, we briefly review three commonly considered architectures: server-based FL, fully decentralized, and semi-decentralized learning, depicted in Fig. \ref{fig:topology}.
	
	\subsubsection{Server-Based FL (Fig. \ref{fig:topology}(a))} 
	The canonical FL algorithm is FedAvg \cite{mcmahan2017communication}, which can be viewed as a generalized version of distributed synchronous SGD with multiple local steps, partial client participation, under data and system heterogeneity. 
	With FedAvg, each device communicates directly with the PS in every communication round after performing a certain number of local SGD steps. Communication costs can be measured, for instance, by the number of participating clients per communication round \cite{chen2022optimal}, or the number of quantization bits per communication link \cite{alistarh2017qsgd, sun2019communication}.
	When the number of participating clients is very large, client sampling/selection is necessary to reduce the communication costs. 
	Furthermore, for large-dimensional models, gradient quantization and compression become necessary to fit transmissions within the available frequency-time resources.
	Both model compression and client sampling contribute to an increased level of distortion in aggregated local updates, causing the global model to oscillate or converge to a biased solution \cite{cho2022towards}. 
	This effect is further exacerbated under non-independent and identically distributed (non-IID) local data distributions and heterogeneous propagation environments \cite{11475389},
	underscoring the challenge of resource allocation in FL under data and channel heterogeneity.
	
	To reduce communication overhead, FedAvg performs multiple local SGD steps between successive aggregation rounds.
	The number of local SGD steps per communication round is another important design parameter in FedAvg that affects the communication costs. Performing multiple local updates reduces communication frequency but gives rise to the \emph{client drift} phenomenon, caused by data heterogeneity across clients \cite{karimireddy2020scaffold, li2019convergence}. In particular, model updates may drift away from the global optimum during local training, degrading convergence performance and final accuracy.
	This requires a careful design on the number of local SGD steps to balance the available communication budget and the convergence performance.
	
	\subsubsection{Decentralized Learning (Fig. \ref{fig:topology}(b))} 
	\label{sec:cost-decentralized}
	In fully decentralized learning systems, each client or agent communicates with its local neighbors via peer-to-peer information exchange. With standard DGD or DSGD, one communication phase (consensus mixing) occurs after every local GD or SGD iteration. More recent studies have considered extended settings with multiple gradient and consensus steps \cite{berahas2021convergence}. Depending on whether unicast or broadcast transmission is considered, the communication costs can be measured by either the number of activated nodes or links per iteration of the algorithm, or by the total number of bits required to achieve a target accuracy \cite{9782148}.
	Meanwhile, implementing decentralized learning algorithms over wireless networks requires medium access control to avoid packet collisions caused by co-channel interference. This introduces additional constraints that determine which set of nodes or links can be active simultaneously. 
	Activating a smaller number of nodes or links in each iteration is equivalent to creating time-varying sparser communication graphs under a given base topology. 
	Reducing communication costs by activating fewer nodes or links typically results in a slower convergence rate, but this effect is non-uniform, as some nodes or links are more critical to network connectivity than others. This observation highlights the importance of node/link scheduling design in balancing convergence speed and communication costs.
	
	\subsubsection{Semi-Decentralized Learning (Fig. \ref{fig:topology}(c))} 
	Recently, semi-decentralized learning has attracted increasing attention. It follows a hierarchical network architecture that combines local consensus and server-assisted model aggregation \cite{9705093,semi-decen, 10843077, rodio2025unified}. In this setting, there are two types of communication links: device-to-device (D2D) information exchange within local clusters and periodic device-to-server (D2S) updates. 
	The strong heterogeneity of communication links across the network hierarchy leads to diverse latency, bandwidth, congestion, and energy characteristics. Moreover, the effective communication cost depends on the communication pattern, update size, transmission strategy, and time-varying channel conditions. Consequently, resource allocation should move beyond minimizing the number of transmissions and instead jointly account for the dynamics of D2D and D2S links and their impact on learning convergence. In particular, D2D aggregation generally incurs lower latency than global aggregation at higher levels of the hierarchy, but provides a more localized view of the network state. This creates a fundamental trade-off between synchronizing D2D cluster models through slower global aggregation, which may rely on outdated models, and relying on faster, more localized D2D updates that better reflect the current state of the system \cite{10304380}.
	
	\subsection{Latency–Reliability Trade-offs in DSGD}
	\label{sec:dsgd-noisy}
	
	Although modern communication systems can provide highly reliable links through channel coding and retransmission mechanisms, they generally incur a significant latency cost, particularly when delivering high-dimensional model updates. Conversely, reducing communication latency generally comes at the expense of sacrificing communication reliability, leading to increased distortion in the exchanged model updates. This latency–reliability trade-off lies at the heart of communication-efficient distributed learning and motivates the design of algorithms that jointly optimize communication and learning performance.
	
	We consider the implementation of DSGD over a wireless network with $\N$ agents as an example. Assuming that each agent broadcasts its local $d$-dimensional model update over orthogonally allocated communication resources, the communication latency per iteration scales as
	$T_{\textnormal{ite}}=\mathcal{O}(\N d)$.
	Combined with the $\mathcal{O}(1/t)$ MSE convergence rate of DSGD (see Sec.~\ref{sec:DSGD}), achieving a target accuracy $\epsilon$ requires an overall communication latency of
	$$
	T_{\textnormal{tot}}=\mathcal{O}\left(\N d/\epsilon\right).
	$$
	This scaling becomes prohibitive in densely deployed networks (large $\N$), such as swarms of UAVs, or when optimizing high-dimensional models (large $d$), as in deep neural networks.
	
	To alleviate this communication bottleneck, two complementary approaches have emerged: model compression targets the high-dimensional regime (large $d$)~\cite{pmlr-v97-koloskova19a}, whereas physical-layer transmission/aggregation techniques improve scalability in the large-network regime (large $\N$)~\cite{10680589}.
	These communication-efficient techniques introduce additional distortion into the exchanged model updates.
	In these cases, DSGD can be modeled as
	\begin{equation}
		\bm x(t+1) = (\W \otimes \I_d)\bm x(t)
		- \alpha_t \mathbf{g}(t)
		+ \boldsymbol{n}(t),
		\label{eq:linear_iteration_noisy}
	\end{equation}
	where $\boldsymbol{n}(t)$ denotes the additive error induced during the information mixing stage of DSGD, arising from protocol-dependent communication impairments, such as \emph{fading channels}, \emph{interference}, \emph{link failures}, and \emph{model compression}.\footnote{Link failures are often modeled through a time-varying mixing matrix; alternatively, they can be represented as additive noise, with the noise term capturing the information lost due to failed links. }
	This model specializes to \eqref{dsgd} when 
	$\boldsymbol{n}(t)=0$.
	The error $n_i(t)$ experienced at node $i$ is commonly modeled as
	\begin{equation}
		\mathbb{E}[{n}_i(t) \,|\, \bm x(t)] = {0}, \quad 
		\mathbb{E}[\|{n}_i(t)\|^2 \,|\, \bm x(t)] \leq \sigcom^2,
		\label{noisecond}
	\end{equation}
	where, similar to Assumption \ref{assump:sfo}, 
	$\sigcom^2$ bounds the variance of communication errors.
	
	Unlike stochastic gradient noise, which is multiplied by the learning stepsize $\alpha_t$, communication noise enters the consensus update independently of $\alpha_t$. Consequently, as the learning stepsize decreases, communication noise eventually becomes the dominant source of error.
	A common mitigation strategy is to introduce an \emph{autoregressive consensus filter} (see, e.g., \cite{10680589,10947567,8786146}), leading to
	\begin{equation}
		\bm x(t+1) = (1-\gamma_t)\bm x(t) + \gamma_t (\W \otimes \I_d) \bm x(t)
		- \alpha_t \mathbf{g}(t)
		+ \gamma_t\boldsymbol{n}(t),
		\label{eq:linear_iteration_filtered}
	\end{equation}
	controlled by the consensus stepsize
	$\gamma_t \in (0,1)$.
	
	With constant stepsizes,
	under Assumptions~\ref{assump:connectivity}-\ref{assump:sfo}, 
	the convergence of DSGD under noisy communications has been characterized in \cite[Sec.~V]{10947567},  yielding
	\begin{align}
		\nonumber
		\limsup_{t \to \infty} \frac{1}{\N}\sum_{i=1}^\N\mathbb{E}\!\left[ \|x_i(t) - x^\star\|^2\right]
		\leq 
		&2\frac{\gamma^2\sigcom^2}{\alpha\mu}
		+ 2\frac{\alpha\sigsgd^2}{\mu}\\&
		+ 2\frac{\alpha^2 \kappa^2 \Delta b}{\gamma^2(1-\lambda_2)^2},
		\label{eq:DSGD_ss_comm}
	\end{align}
	provided that the stepsize $\alpha$ is chosen sufficiently small.
	In contrast to stochastic gradient noise, whose contribution vanishes as $\alpha\to 0$, the communication-induced term scales as $1/\alpha$ when $\gamma$ is fixed. Consequently, reducing the learning stepsize alone does not improve the asymptotic error and may even worsen it.
	Controlling the asymptotic error bound in \eqref{eq:DSGD_ss_comm} therefore requires careful tuning of consensus and learning stepsizes to ensure stability and convergence.
	In particular, by choosing a decreasing learning stepsize as $\alpha_t =\mathcal O(1/t)$ and a decreasing consensus stepsize as $\gamma_t = \mathcal O(t^{-3/4})$, the error to the global optimum decays as \cite{10680589}
	\begin{equation}
		\label{eq:convofDSGDwithnoisycomm}
		\frac{1}{\N}\sum\nolimits_{i=1}^\N \mathbb{E}\!\left[\| {x}_i(t) - {x}^\star \|^2\right] = \mathcal O(1/\sqrt{t}).
	\end{equation}
	after $t$ iterations.
	Although the convergence rate deteriorates from $\mathcal O(1/t)$ to $\mathcal O(1/\sqrt{t})$, communication-efficient implementations substantially reduce the per-iteration latency $T_{\textnormal{ite}}$.
	This highlights a fundamental trade-off between communication resources and optimization performance, which is intrinsic to collaborative ML and sets it apart from classical reliability-centric communication paradigms.
	
	\begin{figure*}[!t]
		\centering
		\subfloat[Noiseless]{
			\includegraphics[width=.32\linewidth]{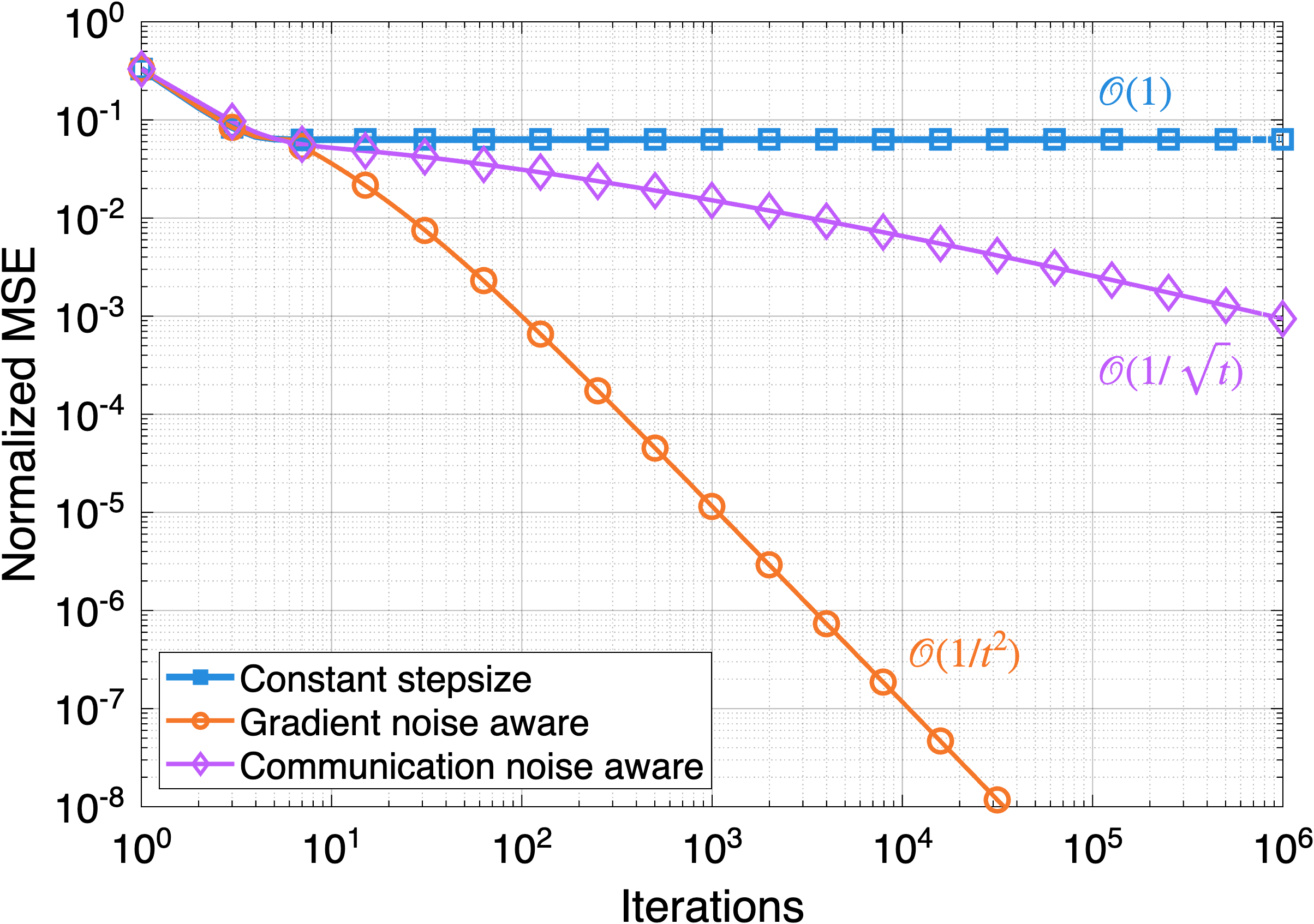}
			\label{fig:const}
		}
		\hfill
		\subfloat[Gradient noise]{
			\includegraphics[width=.3\linewidth]{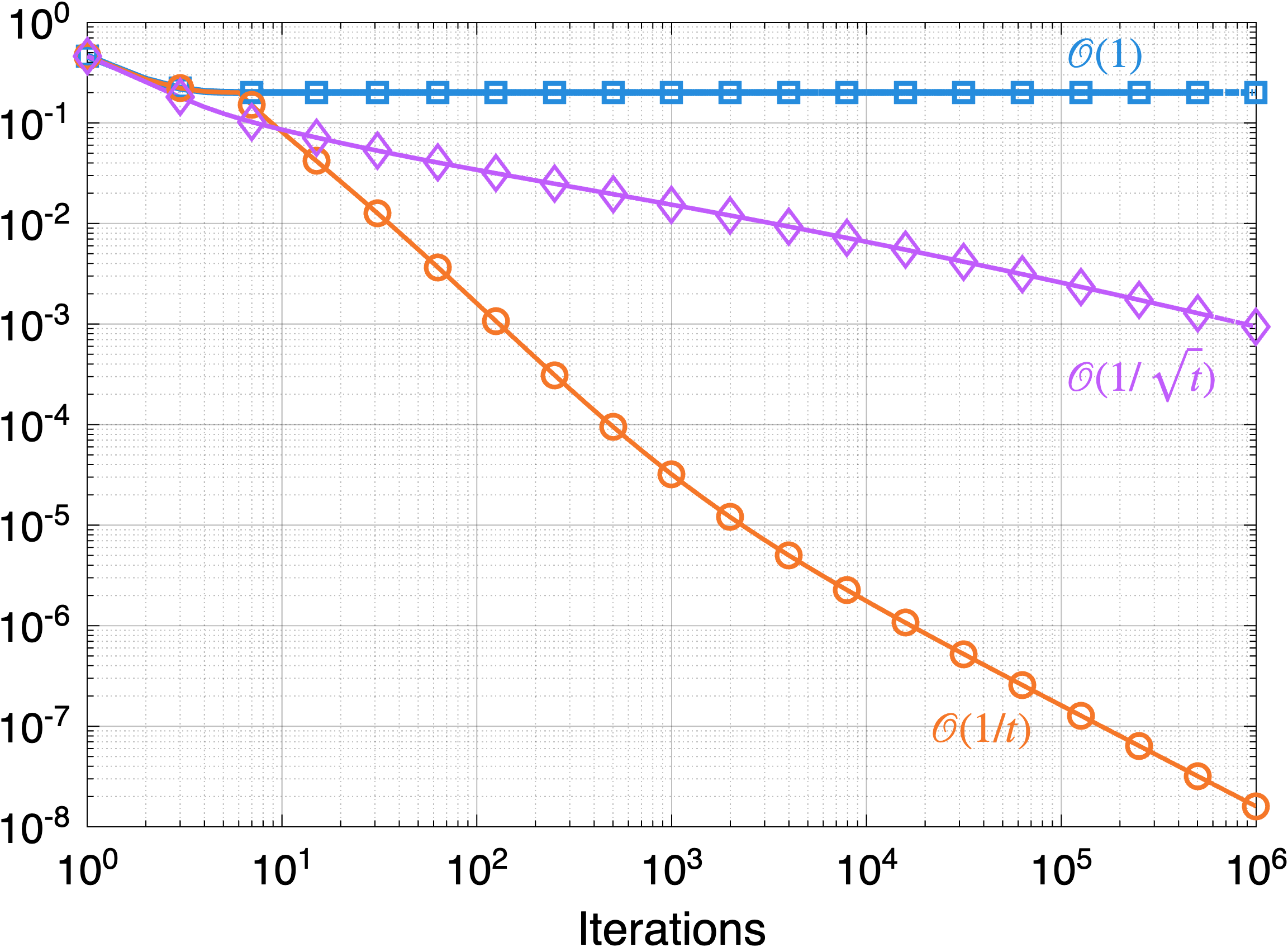}
			\label{fig:grad}
		}
		\hfill
		\subfloat[Gradient and communication noise]{
			\includegraphics[width=.3\linewidth]{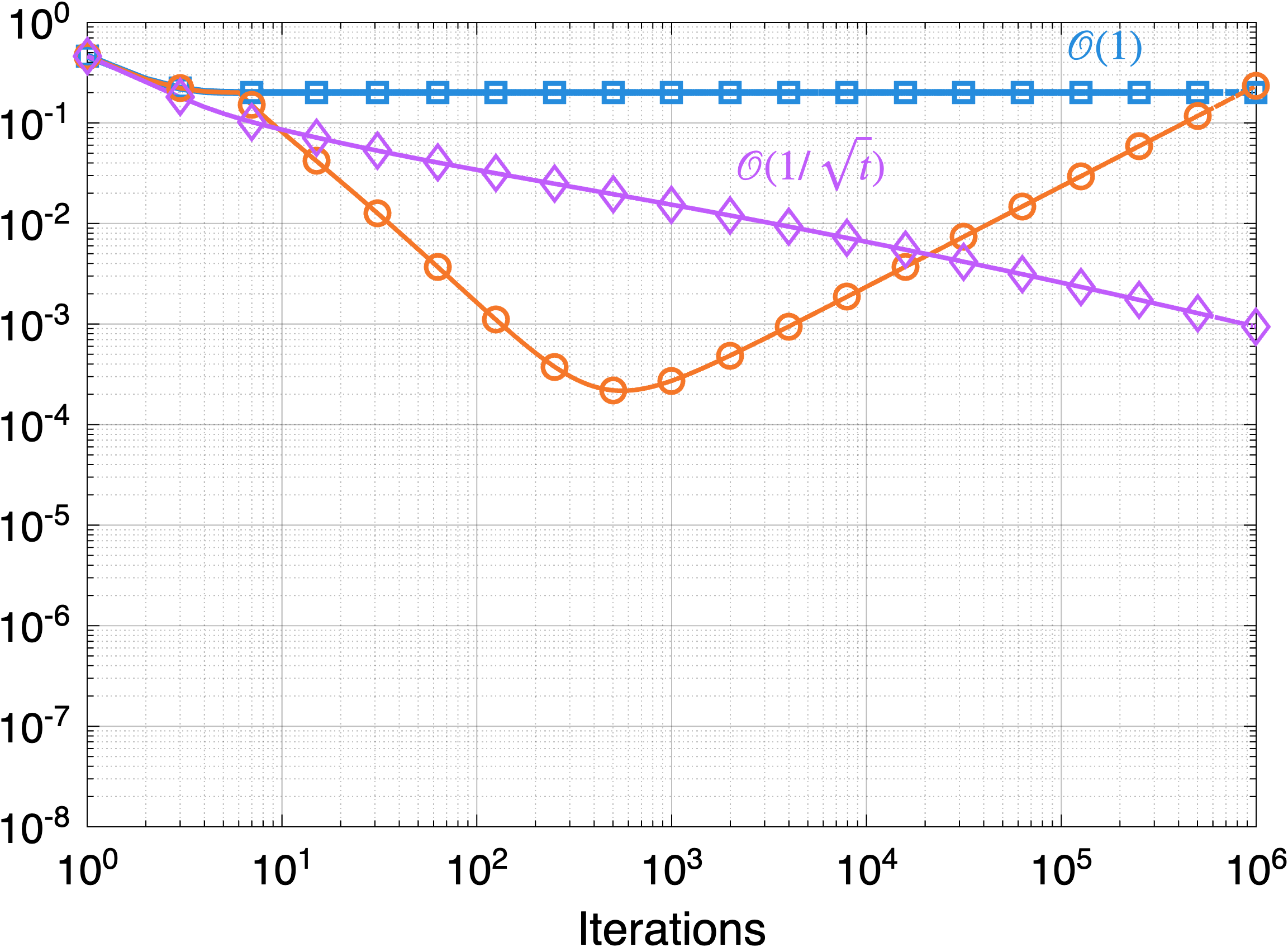}
			\label{fig:comm}
		}
		
		\caption{Comparison of the optimization error under three representative stepsize schedules. Legend shown in the left figure.}
		\label{fig:simulation}
	\end{figure*}
	
	\section{Numerical Example}
	\label{numres}
	As discussed earlier, the convergence behavior of distributed learning algorithms is governed by the interplay between deterministic optimization error, stochastic gradient noise, and communication uncertainty. In this section, we numerically compare these different regimes under representative stepsize sequences, illustrating the theoretical results presented in the previous sections.
	
	We consider a network of $N=50$ agents arranged on a ring, where each agent communicates with its $10$ nearest neighbors on each side (i.e., $20$ neighbors in total). 
	The mixing matrix $\W$ is symmetric and doubly stochastic, with equal weights assigned to all active links. Each agent holds a one-dimensional quadratic objective
	\begin{equation}
		f_i(x)=\frac{\mu_i}{2}\bigl(x-x_i^{\mathrm{loc}}\bigr)^2,
	\end{equation}
	where the curvature parameters $\mu_i$ are independently drawn from the interval $[\mu,L]$ with $\mu=1$ and $L=4$, while the local minimizers $x_i^{\mathrm{loc}}$ are independently generated according to $\mathcal N(0,100)$.
	Gradient noise is modeled as independent additive Gaussian at each agent,
	\(
	\mathcal N(
	\mathbf 0,
	\sigsgd^2\I
	)
	\). Communication noise is modeled as additive Gaussian,
	\(
	\mathcal N(\mathbf 0,\Sigma_{\mathrm{com}})\)
	with covariance
	\(
	\Sigma_{\mathrm{com}}
	=
	\sigma_q^2\W^2.
	\)
	It can be interpreted as the effect of independent quantization errors propagated over the network.\footnote{Quantization errors are generally neither Gaussian nor independent of the state, unless dithering or a high-resolution approximation is employed. Here, we use a Gaussian surrogate to model independent perturbations introduced before network mixing.}
	
	We compare three representative stepsize strategies.
	\begin{list4}
		\item \emph{Constant stepsizes:}
		Both the learning and consensus stepsizes remain fixed,
		\(
		\alpha=\frac{1}{L+\mu},
		\)
		\(\gamma=\frac12,\)
		corresponding to a constant-stepsize DGD/DSGD scheme with a scaled mixing matrix, known to converge linearly to a neighborhood of the optimum.
		\item \emph{Gradient-noise-aware:}
		The consensus stepsize remains fixed,
		\(
		\gamma=\frac12,
		\)
		while the learning stepsize decreases 
		as
		$\alpha_t=\mathcal{O}(1/t)$ 
		following classical stochastic approximation rules \cite{doi:10.1137/070704277}. This strategy mitigates the effect of stochastic gradient noise but does not explicitly account for communication noise.
		\item \emph{Communication-noise-aware:}
		Both the learning and consensus stepsizes decrease over time
		as $\alpha_t=\mathcal O(1/t)$
		and
		$\gamma_t=\mathcal O(1/t^{3/4})$ respectively,
		following the theoretically motivated design discussed in Sec.~IV-C and in~\cite{10680589}. The decreasing consensus stepsize progressively attenuates the accumulation of communication noise while preserving convergence.
	\end{list4}
	
	Fig.~\ref{fig:simulation} compares the MSE (normalized to the initialization error) in three different settings: noiseless optimization (Fig.~\ref{fig:simulation}(\ref{fig:const})), optimization with stochastic gradient noise only (Fig.~\ref{fig:simulation}(\ref{fig:grad})), and optimization with both gradient and communication noise (Fig.~\ref{fig:simulation}(\ref{fig:comm})).
	
	In the noiseless case, the constant-stepsize method converges only to a neighborhood of the optimum due to the steady-state bias of DGD,
	while the gradient-noise-aware stepsize sequence converges to the exact optimum with the theoretical rate $\mathcal{O}(1/t^2)$ (Sec. \ref{sec:DGD}).
	
	When gradient noise is introduced (Fig.~\ref{fig:simulation}(\ref{fig:grad})), the gradient-noise-aware stepsize sequence achieves the asymptotic $\mathcal{O}(1/t)$ decay predicted by the theory (Sec.~\ref{sec:DSGD}), reflecting the effect of stochastic gradient noise. Since communication remains ideal, decreasing the consensus stepsize provides no additional benefit, and the communication-noise-aware stepsize sequence therefore converges more slowly.
	
	Finally, Fig.~\ref{fig:simulation}(\ref{fig:comm}) considers the practical setting where both stochastic gradient and communication noises are present. In this case, the gradient-noise-aware stepsize sequence initially decreases the error rapidly, but communication noise ultimately dominates, leading to performance degradation. In contrast, the communication-noise-aware stepsize sequence progressively decreases both the learning and consensus stepsizes, mitigating the accumulation of communication noise and achieving the asymptotic theoretical rate $\mathcal{O}(1/\sqrt{t})$, discussed in Sec.~\ref{sec:dsgd-noisy}. These results demonstrate that communication impairments fundamentally alter the convergence behavior of distributed optimization algorithms, making communication-aware algorithm design essential for achieving both reliable convergence and communication efficiency.
	
	\section{Communication-Efficient Methods }
	\label{sec:comm-efficiency}
	
	When implementing distributed consensus, optimization, and learning algorithms in large-scale wireless networks where bandwidth, energy, and spatial resources are limited, communication efficiency becomes a central design objective.
	Various communication-efficient methods have been developed and analyzed, including link or node scheduling, data compression (such as quantization, sparsification, low-rank approximation, compressed sensing), and event-triggered communication. 
	In this section, we review commonly considered approaches for improving communication efficiency and emerging techniques that leverage the unique properties of wireless channels. 
	
	\subsection{Communication-Efficient Distributed Consensus}
	
	Earlier research on distributed consensus often models the information exchange as a deterministic process. However, many of the most promising applications of consensus mechanisms, such as multi-robot control and sensor fusion, rely on wireless connectivity in practice. 
	When implementing consensus protocols in wireless networks, the inherent unreliability of the wireless medium (e.g., packet losses, variable latency, and intermittent connectivity) has prompted a shift from deterministic to probabilistic performance guarantees. 
	In particular, the broadcast nature of the wireless medium introduces collisions and co-channel interference, making node/link scheduling and power control two important design elements for achieving fast and reliable information fusion. Moreover, limited bandwidth and energy constraints motivate communication-efficient methods based on compressed updates.
	
	\subsubsection{Node and Link Scheduling}
	Earlier studies assume that all nodes communicate with their neighbors in every iteration. Later on, motivated by the communication and computation constraints in sensor networks, gossip algorithms with randomized pairwise interactions have been extensively studied \cite{dimakis2010gossip, boyd2006randomized, 4497789}.
	This can be viewed as an example of communication scheduling where one link is randomly sampled in each iteration. 
	Subsequent studies have considered various forms of communication scheduling and partial activation to improve communication efficiency, such as event-triggered communication  \cite{dimarogonas2011distributed, seyboth2013event, nowzari2016distributed} and probabilistic link scheduling \cite{antunes2011average, 4536069}.
	Considering the broadcast nature of wireless channels, \cite{probabilistic-broadcast} proposes a broadcast-based scheduling policy with various ways of designing the node transmission probabilities. 
	These works demonstrate that appropriate scheduling mechanisms can significantly reduce communication costs while maintaining convergence to average consensus.
	
	\subsubsection{Power Control}
	In wireless networks with potential link failures, power control directly influences communication reliability and, consequently, the effective network connectivity. Motivated by this observation, several studies have investigated power allocation strategies in distributed consensus.  In \cite{sardellitti2011optimal}, a joint topology control and transmit power optimization problem is formulated with the aim of minimizing the total energy consumption required to achieve consensus. 
	Considering the presence of interference in wireless networks, \cite{vanka2010power} analyzes how transmission power and interference impact both convergence speed and energy costs.
	
	\subsubsection{Compressed Updates}
	As discussed in Sec.~\ref{sec:consensus-noisy}, directly implementing consensus averaging algorithms with quantized state values may lead to persistent steady-state errors.
	Recent work considers error compensation using additional state variables, leading to the CHOCO-Gossip algorithm introduced in \cite{pmlr-v97-koloskova19a}, which achieves convergence to the exact solution even with biased quantization operators.
	For directed network topologies, \cite{9222206} develops a fully decentralized quantized algorithm for designing the mixing weights, coupled with a quantized consensus update rule, and establishes almost sure convergence to the exact average consensus.
	
	In summary, existing studies have shown that power allocation, topology design, transmission scheduling, and compression jointly affect both convergence performance and energy efficiency in wireless consensus systems.
	
	\subsection{Communication-Efficient FL}
	Communication efficiency is essential for scaling FL to large networks \cite{chen2021communication}. Existing approaches can be broadly categorized into several complementary directions, including reducing message sizes through compression, limiting communication links through scheduling, learning-aware resource allocation, and exploiting over-the-air computation (AirComp).
	
	\subsubsection{Model Update Compression}
	A first major line of work develops \textit{gradient or model update compression} techniques that reduce the payload size in each communication round. Communication overhead can be reduced by either decreasing the number of transmitted elements or compressing their digital representation. The former is typically achieved through dimensionality reduction techniques, including sparsification \cite{alistarh2018convergence,adaptive_sparsification}, low-rank approximation \cite{wang2018atomo,vogels2019powersgd}, and subspace projection \cite{park2023regulated}. The latter is commonly accomplished via quantization techniques, including scalar and vector quantization \cite{alistarh2017qsgd, shlezinger2020uveqfed}.
	
	\subsubsection{Client Selection}
	Earlier works on FL have considered various client selection policies under the static channel setting. 
	\cite{chen2022optimal} uses a norm-based importance criterion and derives the optimal participation probabilities for client scheduling. \cite{fu2023client} reviews common client selection strategies, including optimization-based, importance-based, and adaptive sampling approaches. Event-triggered communication is another popular mechanism that reduces communication frequency by avoiding insignificant update transmissions \cite{chen2018LAG}. In time-varying wireless environments, \cite{yang2020scheduling} shows that different device selection and scheduling policies can substantially affect convergence speed. 
	Extending from synchronous training and aggregation, \cite{scheduling-hu} considers asynchronous FL and proposes a joint scheduling and aggregation design that accounts for data similarity, channel conditions, and model staleness.
	
	\subsubsection{Joint Learning-Communication Design}
	In addition to algorithmic approaches, another line of research considers joint learning–communication design that integrates FL with wireless resource management. 
	Representative studies formulate optimization problems that integrate device scheduling, transmit power control, and bandwidth allocation to minimize training loss, latency, or total energy consumption under wireless constraints \cite{Wang2019Adaptive, yang2021energy, shi2021joint, energy-lyapunov-hu,11475389}. The analysis in \cite{chen2022joint} derives explicit expressions linking convergence to wireless resource conditions, providing a theoretical foundation for communication-aware resource allocation.
	Overall, the literature suggests that practical FL in communication-constrained wireless networks benefits from a combined design of learning algorithms, update compression, and resource management. 
	
	\subsubsection{AirComp}
	Traditionally, multi-user wireless communication systems rely on orthogonal resource allocation, ensuring interference-free transmissions by dividing time and frequency resources among devices. Such orthogonal resource allocation becomes inefficient in FL settings with an extremely large number of participating devices, yielding a communication overhead that grows linearly with network size. To overcome this limitation, AirComp has emerged as a promising physical-layer technique for low-latency wireless data aggregation. By exploiting the superposition property of wireless channels, AirComp enables the direct computation of aggregate functions, including arithmetic means, weighted sums, and geometric means, 
	in one shared channel use per scalar component, independently of the number of devices \cite{zhu2021over,ota-chen, ota-survery}. 
	More generally, any nomographic function can be implemented via AirComp through appropriate pre-processing at the transmitter and post-processing at the receiver \cite{goldenbaum2013harnessing}. 
	This approach transforms the communication process from mere data transfer to \emph{in-network computation}, making AirComp particularly well suited for model update aggregation in FL \cite{8870236, 9666041}. 
	Prior work has developed techniques to enhance AirComp-assisted FL, including adaptive learning rate control \cite{9562487, 11079930}, hybrid beamforming \cite{11082012}, retransmissions \cite{otaFL-retrans}, and interference management \cite{9298474}. Recently, AirComp has been extended to other emerging areas such as distributed sensing and inference \cite{10843140, 10186369}.
	Despite its advantages in low latency and high resource efficiency, 
	analog AirComp is sensitive to noise, channel estimation errors, and synchronization mismatches, which may limit its robustness in practical systems. Motivated by these limitations, recent works have proposed digital AirComp methods that integrate quantization, coding, and digital modulation to improve robustness while preserving efficient in-network computation \cite{10893712,10648926, channel-comp, sum-comp}.
	
	Although research in this direction has grown substantially, a comprehensive theoretical framework that explicitly relates channel estimation accuracy and calibration mismatch to per-round aggregation MSE and, ultimately, to end-to-end convergence remains lacking. Moreover, a systematic comparison between digital communication and analog AirComp remains an open research problem that deserves further investigation.
	
	\subsection{Communication-Efficient Decentralized Learning}
	Decentralized learning relies on spatially distributed peer-to-peer information exchange among neighboring nodes. Recent research on communication-efficient decentralized learning focuses on communication scheduling, quantized information exchange, and  AirComp-based aggregation. More recently, non-coherent AirComp has emerged as a scalable alternative that eliminates the need for channel state information, explicit scheduling, and topology-aware coordination.
	
	\subsubsection{Communication Scheduling}
	As discussed in Sec.~\ref{sec:cost-decentralized}, not all nodes or links are equally important in terms of their impact on the network connectivity. In \cite{MATCHA}, a matching-based link scheduling design is proposed, 
	which allows connectivity-critical links to be activated with higher probabilities under a given communication budget. Similar ideas have been explored in \cite{laplacian-sampling,faster-convergence}, where various node or link scheduling policies and mixing matrix optimization designs are developed under different communication models and cost measures. Event-triggered communication is another example of communication scheduling, where information exchange is initiated only when a predefined triggering condition is satisfied, thereby reducing communication overhead while maintaining the desired algorithmic performance
	\cite{9296963}. 
	
	\subsubsection{Quantized Updates}
	Recent studies have considered quantized updates and ways to mitigate the steady-state error caused by quantization noise \cite{9782148,pmlr-v97-koloskova19a, 10155165, differential-nassif-2025}. 
	Adaptive quantization has also been explored, where the quantization resolution increases over time to progressively reduce quantization error. For example, \cite{9782148} proposes a finite-bit quantization framework applicable to a broad class of linearly convergent distributed algorithms, preserving linear convergence while characterizing the communication budget required to achieve a target accuracy.
	
	\subsubsection{AirComp for Decentralized Aggregation}
	Most existing AirComp-based learning frameworks assume star topologies, where multiple clients communicate with a central server (e.g., a base station).
	Recently,  AirComp has been explored for fully decentralized multi-agent consensus and optimization, enabling one-step aggregation without centralized coordination \cite{10680589,9958941,9563232,10506083}.
	Decentralized AirComp-based consensus and optimization algorithms have been analyzed under both single-antenna
	\cite{9563232}
	and multi-antenna \cite{10506083} configurations.
	Several recent works have combined AirComp with decentralized learning by decomposing the network into non-interfering subgraphs using graph coloring \cite{9563232,9322286,9517780}. Within each subgraph, a designated node performs AirComp-based aggregation, followed by a consensus step akin to DSGD. Although effective, these approaches require topology information and channel state information (CSI) feedback, leading to substantial signaling overhead in large networks.
	
	To address these signaling and scalability challenges,
	recent work \cite{10680589} proposes 
	\emph{non-coherent over-the-air aggregation} to achieve fully scalable decentralized learning without scheduling, topology information, or CSI. This approach leverages noisy energy superposition to obtain unbiased consensus estimates using average channel gains as implicit consensus mixing weights, eliminating explicit weight design and CSI acquisition. This framework accommodates general fading and frequency-selective channels, and has been extended to achieve robustness against interference \cite{NCOTA-interference} and to incorporate power control for convergence acceleration \cite{deng2025robust}.
	
	\section{Open Problems and Emerging Directions}
	\label{sec:open}
	This section outlines key open problems and topics of interest for future research, including more realistic assumptions and convergence criteria, fundamental trade-offs, secure and resilient operation, and modern learning scenarios.
	\vspace{-0.2cm}
	\subsection{Mismatch Between Theoretical Assumptions and Reality}
	\label{mismatch}
	\blue{Theoretical analyses of distributed optimization necessarily rely on idealized assumptions regarding the optimization problem, the algorithm, and the underlying system. Although these assumptions enable rigorous convergence guarantees, they often differ significantly from those encountered in practical distributed learning applications. In this section, we discuss several important mismatches between theory and practice, focusing on assumptions on local objective functions, learning-rate design, system-level heterogeneity, and system dynamics.}
	
	\subsubsection{Assumptions on Local Objectives}
	Much of the literature assumes well-behaved objective functions, including strong convexity, Lipschitz-continuous gradients, bounded gradients, or bounded SGD variance.
	These conditions often fail in practice, since typical neural networks are nonconvex and may violate Lipschitz continuity and bounded gradient assumptions. This motivates the development of analytical tools that accommodate broader function classes.
	However, enlarging the function class typically leads to weaker convergence guarantees, which makes the resulting bounds overly conservative when used for resource allocation or parameter tuning.
	\blue{\subsubsection{Adaptive Learning Rates} 
		Most convergence analyses assume carefully designed constant or diminishing stepsize schedules. In practice, however, such schedules are rarely known a priori and often require extensive hyperparameter tuning. Adaptive learning-rate methods alleviate this burden by automatically adjusting the learning rate according to the optimization history~\cite{JMLR:v12:duchi11a,ADAM:2015}. Extending adaptive optimization to federated and decentralized settings remains challenging because learning-rate adaptation must rely on local information while preserving network-wide convergence guarantees. Representative approaches include~\cite{Nedic:2024,Rikos2025DistributedOA}.}
	
	\subsubsection{System-Level Heterogeneity}
	\blue{Practical distributed learning systems are highly heterogeneous. Agents may differ substantially in processing power, energy availability, communication bandwidth, and local data characteristics. In addition, communication delays, intermittent participation and packet losses can lead to asynchronous operation with unpredictable system dynamics. These sources of system-level heterogeneity introduce challenges that are often not captured by classical assumptions and convergence analyses.
		Developing distributed optimization algorithms and analytical frameworks that explicitly account for realistic system heterogeneity remains an active area of research \cite{10665974}.}
	
	\subsubsection{Correlated System Dynamics}
	\blue{Most convergence analyses assume that stochastic gradients, communication noise, or network variations are independent across iterations. 
		These assumptions greatly simplify the analysis by enabling the use of classical stochastic approximation techniques. 
		Practical distributed learning systems, however, exhibit significant temporal and spatial correlations \cite{edin2026exploiting}. Examples include time-correlated wireless fading channels \cite{350282}, bursty packet losses, temporally correlated data \cite{doi:10.1137/20M1355392}, and dynamically evolving environments \cite{cutler2023stochastic}.
		Such correlations may substantially alter the optimization dynamics and invalidate convergence analyses based on independence assumptions. Developing analytical frameworks that explicitly account for correlated stochastic processes therefore remains an active research direction.
	}
	
	\subsection{Convergence Criteria}
	The overwhelming majority of the literature on decentralized optimization uses the MSE (i.e., the expected squared Euclidean distance between the algorithm's iterates and the optimizer in convex optimization, or the expected squared Euclidean norm of the gradient in non-convex optimization) as the primary convergence criterion, providing the number of iterations required to reach a target $\epsilon$-accuracy. While mean-square convergence characterizes the algorithm's average behavior, it does not guarantee convergence along individual sample paths. This motivates the study of stronger notions such as almost-sure convergence, which guarantees convergence with probability one \cite{Nedic10}. Establishing such guarantees typically relies on martingale-based techniques, most notably the Robbins-Siegmund supermartingale convergence theorem \cite{ROBBINS1971233}, together with suitable assumptions on the underlying stochasticity and the stepsize sequence.
	
	\subsection{Interplay between Communication and Computation}
	Distributed learning  relies on repeated local computation and communication among resource-constrained devices, making the joint optimization of communication and computation essential~\cite{nedic2018network}. 
	A central challenge is to allocate communication and computing resources, including local updates, mini-batch sizes, communication rounds, and communication reliability, to achieve a target learning accuracy under latency and energy constraints.
	Despite the extensive literature on communication-efficient techniques, most existing approaches optimize only a subset of the available design parameters while treating the remaining ones as fixed. As a result, they provide only a partial characterization of the underlying communication-computation trade-off. Recent work has investigated unified resource allocation frameworks. For example, \cite{8528465} proposes heuristic rules for balancing communication and computation steps in decentralized optimization, while \cite{bottleneckDGD} formulates decentralized learning under an explicit communication-computation budget and derives adaptive bottleneck-aware scheduling policies.
	Nevertheless, a general theory for jointly optimizing communication, computation, and wireless resources remains largely unexplored. 
	
	\subsection{Privacy, Security, and Resilience}
	\blue{While distributed learning keeps data local and thus provides inherent privacy advantages, it remains vulnerable to various privacy and security threats, including gradient inversion, membership inference, and model poisoning attacks~\cite{trusted-AI}. To protect data confidentiality, two widely adopted approaches are \emph{secure aggregation}, which leverages secure multi-party computation to prevent disclosure of individual model updates~\cite{bonawitz2017practical}, and \emph{differential privacy}, which injects carefully calibrated random noise into local updates~\cite{abadi2016deep}. To enhance robustness against unreliable or malicious participants, Byzantine-resilient aggregation methods~\cite{blanchard2017machine, el2018hidden, chen2017distributed} identify and suppress adversarial updates using robust statistical techniques. Consequently, distributed learning algorithms must operate accurately and reliably under multiple sources of uncertainty, including artificial perturbations introduced for privacy preservation and malicious manipulations.}
	
	\subsection{Modern Topics}
	Our analysis and case studies in this article have focused on single-task distributed learning with static datasets and first-order optimization methods. Extensions to multi-task learning \cite{multi-task}, continual learning \cite{distributed-continual}, 
	zeroth- \cite{YUAN2024111328}
	and second-order methods \cite{dal2024shed} bring many new questions into the communication optimization problem. In multi-task decentralized learning, agents are connected at the topology and task levels, which complicates the effects of communication scheduling on the system performance. With continual learning and streaming datasets, the local objectives become time-varying, which creates another source of dynamics in the importance of model updates from different nodes. 
	Another promising direction is integrating finite-time consensus with communication-aware distributed optimization \cite{finite-time-GT} to accelerate convergence and improve learning efficiency in resource-constrained networks.
	
	\section{Conclusion}
	\label{sec:conclusion}
	In this tutorial, we provided a comprehensive review of the fundamentals of distributed consensus and optimization over networked agents, highlighting the critical role of information mixing and network topology in system performance. We then examined how communication resource constraints affect distributed learning algorithms in practical settings with diverse communication architectures, including federated, decentralized, and semi-decentralized systems. Building on these foundations, we reviewed communication-efficient approaches for distributed learning, including compression, client/link scheduling, learning-aware resource allocation, and joint communication and computation designs such as AirComp. \blue{Beyond summarizing existing works, this tutorial provides a unified framework for understanding multi-agent cooperation and learning under various sources of information uncertainty. We emphasize that communication is not merely an implementation detail, but a fundamental design component of scalable and efficient decentralized AI systems.}
	
	Despite significant progress in related research areas, numerous challenges and open problems remain. These include a deeper understanding of the mismatch between theoretical assumptions and reality, balancing computation, communication, and privacy trade-offs, and developing theoretical guarantees for increasingly complex and stochastic learning environments. 

\end{document}